\documentclass[amsmath,
 amssymb,
 aps,
 nofootinbib,
 preprintnumbers
]{revtex4-1}
\usepackage[utf8]{inputenc}
\usepackage{simplewick}
\usepackage{footnote}
\usepackage{graphicx}
\usepackage{dcolumn}
\usepackage{bm}
\usepackage{stmaryrd}
\usepackage{amsmath}
\usepackage{amssymb}
\usepackage{mathtools}
\usepackage{bbm}
\usepackage{amsfonts}
\usepackage{xcolor}
\usepackage[multiple]{footmisc}
\usepackage{comment}
\usepackage{ulem}
\usepackage{slashed}
\usepackage{multirow}
\usepackage{appendix}
\usepackage{float}
\usepackage{braket}
\usepackage[linktocpage=true,bookmarksnumbered=true,colorlinks=true,plainpages,linkcolor=blue,citecolor=red]{hyperref}

\allowdisplaybreaks

\begin{document}
\preprint{ADP23-28/T1237, MSUHEP-23-031}
\title{Symmetries, Spin-2 Scattering Amplitudes, and Equivalence theorems\\  in Warped Five-Dimensional Gravitational Theories}

\author{R. Sekhar Chivukula$^{a}$}
\author{Joshua A. Gill$^{b}$}
\author{Kirtimaan A. Mohan$^{c}$}
\author{Dipan~Sengupta$^{b}$}
\author{Elizabeth H. Simmons$^{a}$}
\author{Xing Wang$^{a}$}

\affiliation{$^{a}$Department of Physics and Astronomy, University of California, San Diego, 9500 Gilman Drive, La Jolla-92093, USA
}
\affiliation{$^{b}$ARC Centre of Excellence for Dark Matter Particle Physics, Department of Physics, University of Adelaide, South Australia 5005, Australia}
\affiliation{$^{c}$Department of Physics and Astronomy, 
Michigan State University\\
567 Wilson Road, East Lansing, MI-48824, USA
}

\begin{abstract}
Building on work by Hang and He, we show how the residual five-dimensional diffeomorphism symmetries of compactified gravitational theories with a warped extra dimension imply Equivalence theorems which ensure that the scattering amplitudes of helicity-0 and helicity-1 spin-2 Kaluza-Klein states equal (to leading order in scattering energy) those of the corresponding Goldstone bosons present in the `t-Hooft-Feynman gauge. We derive a set of Ward identities that lead to a transparent power-counting of the scattering amplitudes involving spin-2 Kaluza-Klein states. We explicitly calculate these amplitudes in terms of the Goldstone bosons in the Randall-Sundrum model, check the correspondence to previous unitary-gauge computations, and demonstrate the efficacy of `t-Hooft-Feynman gauge for accurately computing amplitudes for scattering of the spin-2 states both among themselves and with matter. Power-counting for the Goldstone boson interactions establishes that the scattering amplitudes grow no faster than ${\mathcal O}(s)$, explaining the origin of the behavior previously shown to arise from intricate cancellations between different contributions to these scattering amplitudes in unitary gauge. We describe how
our results apply to more general warped geometries, including models with a stabilized extra dimension. In an appendix we explicitly identify the symmetry algebra of the residual 5D diffeomorphisms of a Randall-Sundrum extra-dimensional theory.
\end{abstract}

\maketitle


\section{Introduction}

In recent years, motivated by phenomenological studies of massive spin-2 particles, whether in simplified models or in theories of extra dimensions, understanding the behaviour of  scattering amplitudes of massive spin-2 particles has been of increasing interest.
These calculations have implications for cosmological considerations such as dark matter relic density and direct detection~\cite{Garny:2015sjg,Folgado:2019sgz,Lee:2013bua}, and for  phenomenological studies of massive spin-2 resonances at high energy colliders. A key issue is understanding the high energy behaviour of scattering amplitudes with massive spin-2 particles in the external states.

In theories of massive gravity like Fierz-Pauli~\cite{Fierz:1939ix}  and its extensions like dRGT gravity~\cite{deRham:2010kj} in which the mass $m_G$ of the spin-2 is introduced by hand, there is a low-energy strong-coupling scale parametrically lower than the Planck mass ($M_{\rm Pl}$). This follows from the fact that scattering amplitudes of massive gravitons among themselves scale proportional to\footnote{Or at least as fast ~\cite{Arkani-Hamed:1999pwe,Hinterbichler:2011tt,deRham:2014zqa} as ${\mathcal O}(s^{3}/(M_{\rm Pl}^{2}m_{G}^{4})$.} ${\mathcal O}(s^{5}/(M_{\rm Pl}^{2}m_{G}^{8})$ signifying a discontinuity as $m_{G}\to 0$~\cite{Arkani-Hamed:1999pwe,Hinterbichler:2011tt,deRham:2014zqa}. This behavior is an aspect of the van Dam-Veltman-Zakharov (vDVZ) discontinuity~\cite{vanDam:1970vg,Zakharov:1970cc}, a distinctive feature of theories of massive gravity and emerges from the fact that the longitudinal polarization state couples to the trace of the stress-energy tensor. \looseness=-1

In compact extra dimensions, by contrast, massive spin-2 Kaluza-Klein (KK) states~\cite{Kaluza:1921tu,Klein:1926tv} arise from a geometric Higgs mechanism~\cite{Chivukula:2001esy,Lim:2005rc,Lim:2007fy,Lim:2008hi,Chivukula:2022kju,PhysRevD.105.084005,Hang:2022rjp}. In KK theories the massive spin-2 states appear as part of an infinite tower of such states, with the tower starting with a massless spin-2 particle (the graviton), a scalar radion (which is massless in the absence of a mechanism to stabilize the size of the extra dimension), and subsequent spin-2 states with increasing mass spaced in steps proportional to the inverse size of the compact dimension.\looseness=-1

In KK theories the behavior of the scattering amplitudes of these massive spin-2 states is quite different than in Fierz-Pauli and the other theories of massive gravity mentioned above. Individual scattering amplitudes (in unitary gauge) do indeed grow as fast as ${\mathcal O}(s^5/\Lambda^2 M^8_{KK})$, where $\Lambda$ is a mass scale associated with the background geometry, as expected from naive scaling arguments. However there are cancellations between the individual contributions (typically also including cancellations between contributions involving the radion and different intermediate KK tower states) such that the total scattering amplitudes grow no faster than ${\mathcal O}(s/\Lambda^2)$.
 These cancellations have been demonstrated in explicit calculations of the scattering amplitudes in unitary gauge \cite{SekharChivukula:2019qih}. The intricate cancellations involved have been shown to be enforced by a set of sum-rules involving the couplings of the KK states and their masses~\cite{SekharChivukula:2019yul,Bonifacio:2019ioc,Chivukula:2020hvi} (including the radion \cite{Chivukula:2022kju})
 regardless of whether the internal geometry is flat or warped as in the Randall-Sundrum (RS1) model~\cite{Randall:1999ee,Randall:1999vf}. Similar calculations have demonstrated cancellations and revealed sum-rules \cite{Chivukula:2021xod,Chivukula:2022tla} in Goldberger-Wise (GW) models of stabilized extra dimensions \cite{Goldberger:1999uk,Goldberger:1999un}, and in the scattering of KK gravitons with matter \cite{Chivukula:2023sua}.\footnote{See also \cite{deGiorgi:2021xvm} in the case of brane-localized scalar matter.}$^,$\footnote{If massive external spin-2 particles couple to a conserved current, there are no divergences as $m_{G}\to 0$, whether in Fierz-Pauli theory or for KK states as a result of Ward identities \cite{Gill:2023kyz}. }\looseness=-1

We demonstrate in this paper that the cancellations observed between the individual contributions to the massive spin-2 KK scattering amplitudes in unitary gauge are a result of the residual five-dimensional diffeomorphism symmetries \cite{Dolan:1983aa,Lim:2007fy,Lim:2008hi,Chivukula:2022kju} of the compactified KK theory. In particular, the residual diffeomorphism invariance allows one to compute the scattering amplitudes in the analog of `t-Hooft-Feynman gauge rather than unitary gauge. Extending the work of \cite{PhysRevD.105.084005,Hang:2022rjp} which considered helicity-0 scattering for KK states arising from flat extra dimensions (toroidal compactification), we show that the Ward identities of a warped KK gravitational theory in `t-Hooft-Feynman gauge relate the scattering amplitudes of the helicity-0 and helicity-1 states (the states whose scattering amplitudes suffer from the largest potential high-energy growth) to those for the scattering amplitudes of the (unphysical) Goldstone scalar and vector particles present in this gauge. Unlike the massive spin-2 particles, whose helicity-0 and helicity-1 states have polarization vectors which grow with energy in all gauges, naive power-counting of the equivalent Goldstone boson amplitudes manifestly grow no faster than ${\mathcal O}(s/\Lambda^2$), explaining that cancellations observed in previous work arise from residual gauge-invariance.

Specifically, in this work we extend the KK `Gravitational Equivalence theorem' (GRET)  introduced by Hang and He \cite{PhysRevD.105.084005,Hang:2022rjp}, analogous to the familiar Equivalence theorem for massive vector-bosons \cite{Cornwall:1974km,Vayonakis:1976vz,Chanowitz:1985hj,He:1992nga,He:2004zr}, to both the helicity-0 and helicity-1 states of massive spin-2 KK boson scattering. The Ward identities we derive in this work provide a transparent power-counting for the energy dependence of the scattering amplitudes, which proves that the residual terms not accounted for by the GRET grow no faster than ${\mathcal O}(s^0)$. By including the subleading residual terms, we also propose a novel method for computing the scattering amplitudes without large cancellations among the different diagrammatic contributions, and give examples in RS1 which show explicitly how the unitary gauge, 't-Hooft-Feynman gauge, and Goldstone calculations agree. \looseness=-1

The rest of the paper is organized as follows. In Section~\ref{sec:ward}, we introduce 't Hooft-Feynman gauge for the RS1 model and derive the Ward identities for the massive KK gravitons. In section~\ref{sec:longitudinal}, we discuss the scattering of helicity-0 and helicity-1 polarized KK gravitons, and apply the Ward identities to eliminate the apparent bad high energy behavior of the external polarization tensors. In section~\ref{sec:example}, we demonstrate the GRET using two explicit examples: (a) the scattering of two KK bulk scalar into two KK gravitons and (b) the scattering of two KK gravitons into two KK gravitons, and we comment on the connection between our results and the double-copy construction suggested by \cite{PhysRevD.105.084005,Hang:2022rjp}. In section~\ref{sec:exact}, we propose a novel method to compute the exact scattering amplitudes involving longitudinally polarized KK gravitons that is free of large cancellations, and demonstrate its better convergence when only a finite number of intermediate KK states are included, in comparison with the traditional computation in the unitary gauge. We conclude in section~\ref{sec:conclusion} with a discussion of the generality of our results and other questions to be addressed by future work. Appendix~\ref{sec:app_1} outlines our notation, while appendix~\ref{sec:app_2} gives the `t-Hooft-Feynman gauge Feynman rules needed for the computations in section~\ref{sec:exact}. Lastly, in appendix~\ref{sec:app_3} we derive the symmetry algebra of the residual 5D diffeomorphisms of a Randall-Sundrum extra-dimensional theory, extending the results of Duff and Dolan \cite{Dolan:1983aa} for toroidal compactifications.\looseness=-1

\section{RS1 Ward identities for KK gravitons}
\label{sec:ward}

In the 5D RS1 model~\cite{Randall:1999ee,Randall:1999vf}, an orbifolded slice of AdS$_5$, the gravitational fields can be decomposed into towers of KK four-dimensional modes~\cite{Chivukula:2022kju}, \looseness=-1
\begin{eqnarray}
    h_{\mu\nu}(x^\alpha,z) =&& \sum\limits_{n=0}^{\infty}h_{\mu\nu}^{(n)}(x^\alpha)f^{(n)}(z),\label{eq:KK_1u}\\
    A_{\mu}(x^\alpha,z) =&& \sum\limits_{n=1}^{\infty}A_{\mu}^{(n)}(x^\alpha)g^{(n)}(z),\label{eq:KK_2u}\\
    \varphi(x^\alpha,z) =&&~ \sum\limits_{n=0}^{\infty}\varphi^{(n)}(x)k^{(n)}(z)~,\label{eq:KK_3u}
\end{eqnarray}
where $h^{(0)}_{\mu\nu}$ is the massless graviton field, $h^{(n>0)}_{\mu\nu}$ are the massive KK spin-2 fields, $A^{(n>0)}_{\mu}$ are the massive KK vector Goldstone fields, $\varphi^{(0)}$ is the radion field, and $\varphi^{(n>0)}$ are the massive KK scalar Goldstone fields. Here $z_1 \le z \le z_2$ is the internal compact coordinate, $z_{1,2}$ are the locations of the orbifold fixed points, and the mode wavefunctions $f^{(n)}(z)$, $g^{(n)}(z)$, and $k^{(n)}(z)$ (which are respectively even, odd, and even under orbifold parity) are determined by the geometry of the internal space. A brief description of our conventions is given in Appendix~\ref{sec:app_1} and details can be found in \cite{Chivukula:2022kju} and references therein. \looseness=-1

The quadratic terms of Lagragian of the graviton sector are then given by,
\begin{equation}
    \mathcal{L}_2 = \sum_n \left( \frac{1}{2}h^{(n)}_{\mu\nu}\mathcal{D}_h^{\mu\nu\rho\sigma}h^{(n)}_{\rho\sigma} + \frac{1}{2}A^{(n)}_\mu \mathcal{D}_A^{\mu\nu}A^{(n)}_\nu + \frac{1}{2}\varphi^{(n)} D_\varphi \varphi^{(n)}\right).
\end{equation}
Crucially~\cite{Lim:2007fy,Lim:2008hi}, the wave equations for these modes of different spin are related by a pair of $N=2$ quantum-mechanical SUSY symmetries that enforce the degeneracy of the non-zero mass modes of these different spins, a situation that also holds in the case of a stabilized extra dimension \cite{Chivukula:2022kju}, and hence the inverse propagators are given by\looseness=-1
\begin{eqnarray}
    \mathcal{D}_h^{\mu\nu\rho\sigma} &=& \frac{1}{2}\left(\eta^{\mu\rho}\eta^{\nu\sigma} + \eta^{\mu\sigma}\eta^{\nu\rho} - \eta^{\mu\nu}\eta^{\rho\sigma}\right)(-\square - m_n^2),\\
    \mathcal{D}_A^{\mu\nu} &=& -\eta^{\mu\nu}(-\square - m_n^2),\\
    \mathcal{D}_\varphi &=& -\square - m_n^2.
\end{eqnarray}

In addition, the degeneracy of these different modes allows one to adopt a 't Hooft-Feynman gauge for gravity~\cite{Lim:2007fy,Lim:2008hi} with a warped internal dimension, ( here $h^{(n)} \equiv h^{{(n)}\mu}{}_{\mu}$)
\begin{eqnarray}
    \mathcal{L}_{\rm GF} &=& \sum_nF^{(n)\mu} F^{(n)}_\mu-F^{(n)}_5F^{(n)}_5, \label{eq:GF}\\
    F^{(n)}_\mu &=& -\left(\partial^\nu h^{(n)}_{\mu\nu} - \dfrac{1}{2}\partial_\mu h^{(n)} + \dfrac{1}{\sqrt{2}}m_nA^{(n)}_\mu\right), \label{eq:Fmu} \\
    F^{(n)}_5 &=& -\left(\dfrac{1}{2}m_n h^{(n)} - \dfrac{1}{\sqrt{2}}\partial^\mu A^{(n)}_\mu +\sqrt{\dfrac{3}{2}}m_n\varphi^{(n)}\right).
    \label{eq:F5}
\end{eqnarray}
From the gauge fixing condition, one can derive the Ward identities~\cite{PhysRevD.105.084005,He:1993yd} for the time-ordered matrix elements
\begin{equation}
    \braket{\mathbf{T} F^{(n)}_\mu(x) \Phi} = \braket{\mathbf{T} F^{(n)}_5(x) \Phi} = 0,
\end{equation}
where $\Phi$ denotes any other on-shell physical fields after the
Lehmann-Symanzik-Zimmermann (LSZ) amputation. Plugging in the gauge fixing condition in Eqs.~(\ref{eq:Fmu}) and (\ref{eq:F5}), we have the following identities for the time-ordered Green's functions
\begin{eqnarray}
    &&\braket{\mathbf{T} \left(\partial^\nu(h^{(n)}_{\mu\nu} - \dfrac{1}{2}\eta_{\mu\nu} h^{(n)} ) + \dfrac{1}{\sqrt{2}}m_nA^{(n)}_\mu\right) \Phi} = 0,\\
    &&\braket{\mathbf{T} \left(\dfrac{1}{2}m_n h^{(n)} - \dfrac{1}{\sqrt{2}}\partial^\mu A^{(n)}_\mu +\sqrt{\dfrac{3}{2}}m_n\varphi^{(n)}\right) \Phi} = 0.
\end{eqnarray}
Because of the mass degeneracy of $h^n_{\mu\nu}$, $A^n_\mu$ and $\varphi^n$, we can amputate these external states at the same time by multiplying by $(-\square-m_n^2)$.
\begin{figure}
    \centering
    \includegraphics[width = 0.23\textwidth]{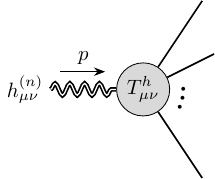} \hspace{.4cm}
    \includegraphics[width = 0.23\textwidth]{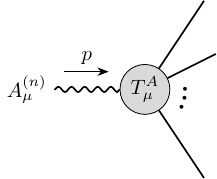} \hspace{.4cm}
    \includegraphics[width = 0.23\textwidth]{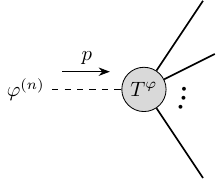}
    \caption{Schematic Feynman diagrams involving one external KK graviton, one KK vector Goldstone boson, or one KK scalar Goldstone boson.}
    \label{fig:feyn1}
\end{figure}

Now consider the processes shown in Fig.~\ref{fig:feyn1}, whose scattering amplitudes, $\mathcal{M}$, can be written, respectively, as
\begin{equation}
    \mathcal{M}^h = \epsilon^{\mu\nu}(p)T^h_{\mu\nu},\qquad\mathcal{M}^A = \epsilon^{\mu}(p)T^A_{\mu},\qquad\mathcal{M}^\varphi = T^\varphi,
\end{equation}
where the $\epsilon^{\mu\nu}$ and $\epsilon^\mu$ are the polarization vectors of the external spin-2 and spin-1 particles.
Note that the sub-amplitudes $T^h_{\mu\nu}$, $T^A_{\mu}$ and $T^\varphi$ are related to the corresponding Green's functions by LSZ amputation,
\begin{eqnarray}
    T^h_{\mu\nu} &=& \mathcal{N} \int d^4x~e^{ipx}~D^h_{\mu\nu\rho\sigma} \braket{\mathbf{T} ~h^{(n)}_{\rho\sigma}(x) \Phi} \nonumber\\
    &=&  \mathcal{N} \int d^4x~e^{ipx}~(-\square - m_n^2) \braket{\mathbf{T} \left(h^{(n)}_{\mu\nu}(x) - \dfrac{1}{2}\eta_{\mu\nu}h^{(n)}(x)\right) \Phi}\\
    T^A_{\mu} &=& \mathcal{N} \int d^4x~e^{ipx}~D^A_{\mu\nu} \braket{\mathbf{T} ~A^{(n)}_{\nu}(x) \Phi} = \mathcal{N} \int d^4x~e^{ipx}~(-\square - m_n^2) \braket{\mathbf{T} \left(-A^{(n)}_{\mu}(x)\right) \Phi}\\
    T^\varphi &=& \mathcal{N} \int d^4x~e^{ipx}~D^\varphi \braket{\mathbf{T} ~\varphi^{(n)}(x) \Phi} = \mathcal{N} \int d^4x~e^{ipx}~(-\square - m_n^2) \braket{\mathbf{T}~\varphi^{(n)}(x) \Phi}
\end{eqnarray}

Assuming the momentum is incoming, one derives the following Ward identities,
\begin{eqnarray}
    \dfrac{i}{2}p^\nu\left(T^h_{\mu\nu} +T^h_{\nu\mu}\right) - \dfrac{1}{\sqrt{2}}m_n T^A_\mu = 0, \label{eq:ward1}\\
    -\frac{1}{2}m_n T^{h\mu}_{\mu} + \dfrac{i}{\sqrt{2}}p^\mu T^A_\mu + \sqrt{\frac{3}{2}}m_nT^\varphi = 0. \label{eq:ward2}
\end{eqnarray}
Note that the derivation of the Ward identities above only relies on the gauge fixing conditions in Eqs.~(\ref{eq:Fmu}) and (\ref{eq:F5}), and will work in any geometry in which these gauge-fixing conditions can be applied. In particular, the same conditions apply in toroidal compactifications \cite{PhysRevD.105.084005, Hang:2022rjp}, and we discuss the generalization to the GW model in Sec.~\ref{sec:example}.
We note that the identity will also work  if the KK graviton is off-shell.

We use these Ward identities to formulate the Goldstone boson Equivalence theorems in the next section.

\section{The Goldstone boson Equivalence theorem in RS1}
\label{sec:longitudinal}

In this section we will use the Ward Identities in Eqs.~(\ref{eq:ward1}) and (\ref{eq:ward2}) to relate amplitudes with one or more helicity-0 or helicity-1 external states with the corresponding Goldstone boson amplitudes in the RS1 model. For the longitudinally polarized (helicity-0) KK graviton external state, the polarization tensor can be expressed using two spin-1 polarization vectors,
\begin{equation}
    \epsilon_0^{\mu\nu} = \dfrac{1}{\sqrt{6}}\left(\epsilon_+^\mu\epsilon_-^\nu + \epsilon_-^\mu\epsilon_+^\nu + 2\epsilon_0^\mu\epsilon_0^\nu\right),
\end{equation}
where the polarization vectors, for momentum with polar angle $\theta$ and azimuthal angle $\phi$, are defined as
\begin{eqnarray}
    \epsilon_\pm^\mu &=& \dfrac{1}{\sqrt{2}}\left(0, \mp\cos\theta\cos\phi+i\sin\phi,\mp\cos\theta\sin\phi-i\cos\phi,\pm\sin\theta\right)^{\rm T},\\
    \epsilon_0^\mu &=& \dfrac{1}{m}\left(\sqrt{E^2-m^2}, E\sin\theta\cos\phi, E\sin\theta\sin\phi,E\cos\theta\right)^{\rm T}.
\end{eqnarray}
The polarization vectors have the energy dependency, when $E\gg m$,
\begin{equation}
    \epsilon_\pm^\mu \sim \mathcal{O}(1),\qquad \epsilon_0^\mu \sim \mathcal{O}(E/m).
\end{equation}
Thus, the longitudinal polarization tensor depends on the energy quadratically at high-energies,
\begin{equation}
    \epsilon_0^{\mu\nu}\sim\mathcal{O}(E^2/m^2),
\end{equation}
leading to large individual contributions when computing the longitudinal KK graviton scattering amplitude in unitary gauge. We show below how to rewrite these polarization vectors such that the Ward identities can be applied, leading to amplitudes with no bad high-energy behavior.\looseness=-1

We begin by re-expressing the amplitudes involving helicity-0 external states using the Ward Identities. Note that one can use the the polarization sum for spin-1 polarization vectors,
\begin{equation}
    \epsilon_\mp^\mu = -\left(\epsilon_\pm^\mu\right)^*,\qquad \sum_{\lambda=\pm,0}\epsilon_\lambda^\mu \epsilon_\lambda^{\nu*} = -\eta_{\mu\nu}+\dfrac{p^\mu p^\nu}{m^2},
\end{equation}
to rewrite the longitudinal polarization tensor as
\begin{equation}
    \epsilon_0^{\mu\nu} = \dfrac{1}{\sqrt{6}}\left(\eta^{\mu\nu} -\dfrac{p^\mu p^\nu}{m^2}  + 3\epsilon_0^\mu\epsilon_0^\nu\right).
\end{equation}
The potentially bad high energy behavior from $\epsilon_0^\mu$ can be isolated by introducing~\cite{Wulzer:2013mza, Chen:2016wkt, Chen:2022gxv}
\begin{equation}
    \epsilon_0^\mu \equiv \frac{p^\mu}{m} + \tilde{\epsilon}_0^\mu,\qquad \text{where } \tilde{\epsilon}_0^\mu \equiv -\frac{m}{E+|\mathbf{p}|} (1, - \mathbf{p}/|\mathbf{p}|) \sim \mathcal{O}(m/E).
    \label{eq:decomposition}
\end{equation}
Thus one can rewrite the longitudinal polarization tensor as 
\begin{equation}
    \epsilon_0^{\mu\nu} = \tilde{\epsilon}_0^{\mu\nu} + \dfrac{1}{\sqrt{6}}\left(\eta^{\mu\nu} + 2\,\dfrac{p^\mu p^\nu}{m^2} + 3\,\frac{p^\mu\tilde{\epsilon}_0^\nu + p^\nu\tilde{\epsilon}_0^\mu}{m}\right),
\end{equation}
where
\begin{equation}
    \tilde{\epsilon}_0^{\mu\nu} \equiv \sqrt{\frac{3}{2}}\tilde{\epsilon}_0^\mu\tilde{\epsilon}_0^\nu \sim\mathcal{O}\left(\frac{m^2}{E^2}\right),
\end{equation}
thereby expressing the external longitudinal polarization tensor in terms of external momenta and sub-leading terms.

Using the Ward identities in Eqs.~(\ref{eq:ward1}) and (\ref{eq:ward2}), we see that
\begin{eqnarray}
    T^h_{\mu\nu}\frac{p^\mu p^\nu}{m_n^2} &=& -\frac{i}{\sqrt{2}}\ T^A_\mu\frac{p^\mu}{m_n} = -\frac{1}{2} \ T^{h\mu}_{\mu} + \sqrt{\frac{3}{2}}\ T^\varphi,\\
    T^h_{\mu\nu}\frac{p^\mu\tilde{\epsilon}_0^\nu + p^\nu\tilde{\epsilon}_0^\mu}{m} &=& - i \sqrt{2} \ T^A_\mu\tilde{\epsilon}_0^\mu.
\end{eqnarray}
Therefore, 
\begin{equation}
    T^h_{\mu\nu}\left(\eta_{\mu\nu}+2\dfrac{p^\mu p^\nu}{m^2}\right) = \sqrt{6}\,T^\varphi~,
\end{equation}
and the longitudinal scattering amplitude can be expressed as 
\begin{equation}
    T^h_{\mu\nu}\epsilon_0^{\mu\nu} = T^\varphi - i\sqrt{3}\,T^A_\mu\tilde{\epsilon}_0^\mu + T^h_{\mu\nu}\tilde{\epsilon}_0^{\mu\nu} .
    \label{eq:ward_A}
\end{equation}
Note that there is no bad high energy behavior coming from the external polarization tensors or vectors on the right hand side of Eq.~(\ref{eq:ward_A}); the second and third terms on the right hand side of Eq.~(\ref{eq:ward_A}) are relatively suppressed due to the fact that $\tilde{\epsilon}_0^\mu\sim\mathcal{O}(m/E)$ and $\tilde{\epsilon}_0^{\mu\nu}\sim\mathcal{O}(m^2/E^2)$. 

This expression confirms and extends the Goldstone boson Equivalence theorem given by Hang and He for longitudinal KK graviton scattering~\cite{PhysRevD.105.084005,Hang:2022rjp}, namely that the scattering amplitude of the longitudinally polarized KK gravitons equals that of the scalar KK Goldstone boson in the high energy limit,
\begin{equation}
    T^h_{\mu\nu}\epsilon_0^{\mu\nu} = T^\varphi +\mathcal{O}(s^0).
    \label{eq:ge_1}
\end{equation}
Furthermore, our derivation of Eq.~(\ref{eq:ward_A}) demonstrates that the  Equivalence theorem is valid for a warped internal space and gives an explicit expression for the residual terms not captured by the leading-order expression.

Similarly, using the definitions of the helicity $\pm 1$ polarization tensors
\begin{equation}
\epsilon^{\mu\nu}_{\pm 1} = \dfrac{1}{\sqrt{2}}\left( \epsilon^\mu_{\pm }\epsilon^\nu_0 + \epsilon^\mu_0 \epsilon^\nu_{\pm }\right)~,    
\end{equation}
and using the decomposition of $\epsilon^\mu_0$ given in Eq.(\ref{eq:decomposition}) and applying the Ward identities, one finds the following identities for helicity $\pm 1$ states,
\begin{equation}
    T^h_{\mu\nu}\epsilon_{\pm 1}^{\mu\nu} = - i T^A_\mu\epsilon_\pm^\mu + T^h_{\mu\nu}\tilde{\epsilon}_{\pm 1}^{\mu\nu},
    \label{eq:ward_B}
\end{equation}
where
\begin{equation}
    \tilde{\epsilon}_{\pm 1}^{\mu\nu} \equiv \dfrac{1}{\sqrt{2}}\left(\epsilon_\pm^\mu\tilde{\epsilon}_0^\nu + \tilde{\epsilon}_0^\mu\epsilon_\pm^\nu\right) \sim\mathcal{O}\left(\frac{m}{E}\right).
\end{equation}
Therefore, the Goldstone boson Equivalence theorem for the helicity $\pm 1$ states is: the scattering amplitude of the KK gravitons with helicities $\pm 1$ equals that of the vector KK Goldstone boson in the high energy limit up to a overall phase,
\begin{equation}
    T^h_{\mu\nu}\epsilon_{\pm 1}^{\mu\nu} = - i T^A_\mu\epsilon_\pm^\mu + \mathcal{O}(s^0).
\end{equation}

While we have derived the above identities for one external KK graviton, one can easily generalize it to the case of multiple external KK gravitons, by examining
\begin{equation}
    \braket{\mathbf{T} F^{(n_1)}_{\mu/5}(x)F^{(n_2)}_{\nu/5}(x)\cdots \Phi} =  0.
\end{equation}
By neglecting the subleading terms, we arrive at the Goldstone Equivalence theorem for the helicity-0 KK gravitons,
\begin{equation}
    \mathcal{M}\left[h_L^{(n_1)}h_L^{(n_2)}\cdots\right] = \mathcal{M}\left[\varphi^{(n_1)}\varphi^{(n_2)}\cdots\right] + \mathcal{O}(s^0),
    \label{eq:helicity-0}
\end{equation}
and for the helicity $\pm1$ KK gravitons,
\begin{equation}
    \mathcal{M}\left[h_{\pm1}^{(n_1)}h_{\pm1}^{(n_2)}\cdots\right] = (-i)^{N_{\rm in}}(i)^{N_{\rm out}}\mathcal{M}\left[A_\pm^{(n_1)}A_\pm^{(n_2)}\cdots\right] + \mathcal{O}(s^0),
    \label{eq:helicity-one}
\end{equation}
where $N_{\rm in}$ ($N_{\rm out}$) is the number of incoming (outgoing) helicity $\pm1$ KK graviton states.

We also note that one can organize the above results into a more compact way by introducing 5D polarization tensors as
\begin{eqnarray}
    \tilde{\epsilon}_0^{MN} = \begin{pmatrix}
        \tilde{\epsilon}_0^{\mu\nu} - \sqrt{\dfrac{1}{6}}\eta^{\mu\nu} & i\sqrt{\dfrac{3}{2}}\tilde{\epsilon}_0^\mu \\
        i\sqrt{\dfrac{3}{2}}\tilde{\epsilon}_0^\nu & -\sqrt{\dfrac{2}{3}}
    \end{pmatrix}, \quad 
    \tilde{\epsilon}_{\pm 1}^{MN} &=& \begin{pmatrix}
        \tilde{\epsilon}_{\pm 1}^{\mu\nu} & \dfrac{i}{\sqrt{2}}\epsilon_\pm^\mu \\
        \dfrac{i}{\sqrt{2}}\epsilon_\pm^\nu & 0
        \end{pmatrix}, \quad
        \tilde{\epsilon}_{\pm 2}^{MN} = \begin{pmatrix}
        \epsilon_{\pm 2}^{\mu\nu} & 0 \\
        0 & 0
        \end{pmatrix},
\end{eqnarray}
and 
\begin{equation}
    T_{MN} = \begin{pmatrix}
        T^h_{\mu\nu} & \ \ \ -\dfrac{1}{\sqrt{2}}T^A_\mu\\
        -\dfrac{1}{\sqrt{2}}T^A_\nu & \ \ \ -\dfrac{1}{2}T^{h\mu}_\mu - \sqrt{\dfrac{3}{2}}T^\varphi
    \end{pmatrix} \label{eq:Tmat},
\end{equation}
such that the above identities relating amplitudes can be written as,
\begin{equation}
    T^h_{\mu\nu}\epsilon_{\lambda}^{\mu\nu} = T_{MN}\tilde{\epsilon}_{\lambda}^{MN}.
\end{equation}

\begin{figure}
    \centering
    \includegraphics[width = 0.23\textwidth]{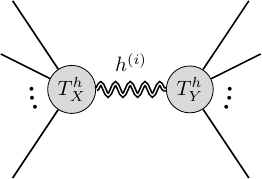}\hspace{0.4cm}
    \includegraphics[width = 0.23\textwidth]{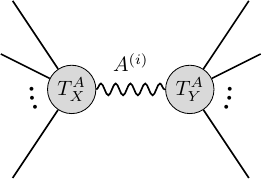}\hspace{0.4cm}
    \includegraphics[width = 0.23\textwidth]{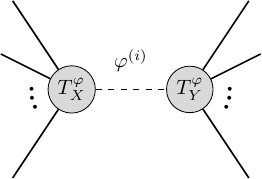}
    \caption{Schematic Feynman diagrams involving one internal KK graviton, one KK vector Goldstone boson, or one KK scalar Goldstone boson.}
    \label{fig:feyn2}
\end{figure}

Correspondingly, we can parametrize the scattering matrix $T_{MN}$ as in Eq.~(\ref{eq:Tmat}) so that the internal propagators can be also written as a 5D graviton propagator that has the tensor structure of a 5D massless graviton. For a scattering shown in Fig.~\ref{fig:feyn2}, the amplitude can be written as 
\begin{equation}
    \mathcal{M} = T^h_{X\,\mu\nu} \mathcal{P}_h^{\mu\nu\rho\sigma} T^h_{Y\,\rho\sigma} + T^A_{X\,\mu} \mathcal{P}_A^{\mu\nu} T^A_{Y\,\nu} + T^\varphi_X \mathcal{P}_\varphi T^\varphi_Y \equiv \hat{T}_{X\, MN} \mathcal{P}^{MNRS} \hat{T}_{Y\, RS},
\end{equation}
where the propagators are give by
\begin{eqnarray}
    &&\mathcal{P}_h^{\mu\nu\rho\sigma} = \frac{i}{p^2-m_n^2}\frac{1}{2}\left( \eta^{\mu\rho}\eta^{\nu\sigma} + \eta^{\mu\sigma}\eta^{\nu\rho} - \eta^{\mu\nu}\eta^{\rho\sigma} \right),\quad \mathcal{P}_A^{\mu\nu} = \frac{-i\eta^{\mu\nu}}{p^2-m_n^2},\quad\mathcal{P}_\varphi = \frac{i}{p^2-m_n^2},\\
    &&\mathcal{P}^{MNRS} = \frac{i}{p^2-m_n^2}\frac{1}{2}\left( \eta^{MR}\eta^{NS} + \eta^{MS}\eta^{NR} - \dfrac{2}{3}\eta^{MN}\eta^{RS} \right).
\end{eqnarray}

In this section, combining the results of \cite{Lim:2007fy,Lim:2008hi} for the RS1 model, we have extended the analysis of \cite{PhysRevD.105.084005,Hang:2022rjp} to establish  the gravitational Goldstone boson Equivalence theorem
for the scattering amplitudes in the compactified RS1 model (see Sec.~\ref{sec:example} for a brief discussion of the GW model).
To leading order, the Goldstone boson Equivalence theorem relates the scattering of helicity-0 and helicity-1 KK gravitons to that of the Goldstone bosons present in `t-Hooft-Feynman gauge, Eqs.~(\ref{eq:helicity-0}) and (\ref{eq:helicity-one}). 
The analysis of the RS1 model in a `t-Hooft-Feynman gauge is only possible because of residual 5D diffeomorphism invariance of the theory, which can be formally described by the algebra given in appendix~\ref{sec:app_3}. Power-counting of the Goldstone boson amplitudes in `t-Hooft-Feynman gauge demonstrates that the scattering amplitudes of KK-gravitons among themselves or with matter can grow no faster than ${\mathcal O}(s)$, explaining the cancellations observed in the unitary gauge calculations of 
\cite{SekharChivukula:2019qih,SekharChivukula:2019yul,Bonifacio:2019ioc,Chivukula:2020hvi}. It is important to remember that the Goldstone boson Equivalence theorem relates the scattering of KK gravitons to that of the Goldstone bosons only to leading order,  $\mathcal{O}(s)$. In the case of vanishing scattering amplitudes at $\mathcal{O}(s)$ due to helicity selection rules, one would have to include the subleading terms in Eqs.~(\ref{eq:ward_A}) and (\ref{eq:ward_B}).

To use these results to compute scattering amplitudes, one must construct the couplings of the Goldstone bosons in `t-Hooft-Feynman gauge. We illustrate this in the next section in RS1, checking that the results agree to leading order with previous unitary-gauge computations. As we explain more completely in Sec.~\ref{sec:conclusion}, however, although the form of the Equivalence theorem will remain the same in other warped geometries, the computation of the Goldstone boson matrix elements will depend on the details of the model.

\section{Applying the Equivalence theorem: Two Examples}
\label{sec:example}

In this section, we apply the Goldstone boson Equivalence theorem in RS1 to the scattering of two longitudinally polarized KK gravitons into a pair of KK scalars, and to the elastic scattering of the longitudinally polarized KK gravitons. We show that, to leading order in energy, the scattering amplitude involving helicity-0 spin-2 particles \cite{Chivukula:2020hvi,Chivukula:2023sua} equals the `t-Hooft-Feynman gauge amplitude for the scalar Goldstone boson, per Eq.~(\ref{eq:ge_1}). We would like to emphasize that, while we choose the RS1 model for our examples in this paper, the form of the Equivalence theorem is generic for other warped background geometries, such as in a GW model \cite{Goldberger:1999uk,Goldberger:1999un}, though the interactions among the Goldstone bosons will differ from those evaluated here in RS1.

\subsection{Scattering of two KK bulk scalars into two helicity-0 KK gravitons}
For the first example, we consider the scattering of two KK bulk scalar into two longitudinal KK gravitons,
\begin{equation}
    S^{(n_1)}S^{(n_2)} \rightarrow h^{(n_3)}_{L}h^{(n_4)}_{L}.
\end{equation}
According to the Goldstone boson Equivalence theorem, one should expect
\begin{equation}
    \mathcal{M}\left[S^{(n_1)}S^{(n_2)} \rightarrow h^{(n_3)}_{L}h^{(n_4)}_{L}\right] = \mathcal{M}\left[S^{(n_1)}S^{(n_2)} \rightarrow \varphi^{(n_3)}\varphi^{(n_4)}\right] + \mathcal{O}(s^0).
\end{equation}
We take the matter Lagrangian for a real bulk scalar $S$ with a mass $M_{S}$ to be (the metric $G^{MN}$ is defined in appendix~\ref{sec:app_1})
\begin{equation}
    \mathcal{L}_{m}  = \sqrt{G}\left(\dfrac{1}{2}G^{MN}\partial_M S\partial_N S - \dfrac{1}{2}M_{S}^2 S^2\right),
\end{equation}
subject to the boundary conditions,\footnote{For simplicity, we consider a model with no bulk potential or brane-localized scalar interactions.}
\begin{equation}
    \partial_z S = 0\quad{\rm at }~z=z_{1,2}.
\end{equation}
In the above expression $\sqrt{G}$ denotes the determinant of the 5D  metric.
Following the notation in Ref.~\cite{Chivukula:2023sua}, we decompose the bulk scalar field into KK modes,
\begin{equation}
    S(x^\alpha,z) = \sum_{n=0}^\infty S^{(n)}(x^\alpha)f^{(n)}_S(z),
\end{equation}
where $f^{(n)}_S$ are the eigenfunctions of the mode equation
\begin{equation}
    \left[\left(-\partial_z-3A'(z)\right)\partial_z + M_S^2e^{2A(z)}\right]f^{(n)}_S = m_{S,n}^2f^{(n)}_S,
\end{equation}
where $A(z)$ is the warp factor in the conformal coordinate line-element (see appendix \ref{sec:app_1}).

\begin{figure}
    \centering
    \includegraphics[width = 0.24\textwidth]{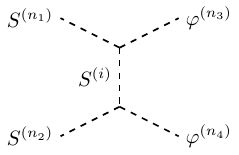}
    \includegraphics[width = 0.24\textwidth]{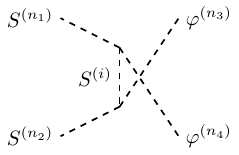}
    \includegraphics[width = 0.24\textwidth]{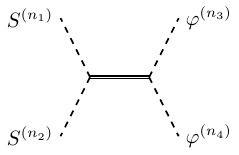}
    \includegraphics[width = 0.24\textwidth]{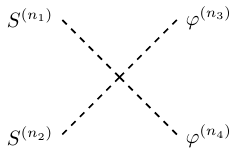}
    \caption{The Feynman diagrams for $S^{(n_1)}S^{(n_2)} \rightarrow \varphi^{(n_3)}\varphi^{(n_4)}$, where we use the double line to indicate all three possible intermediate states $h^{(i)}$, $A^{(i)}$, and $\varphi^{(i)}$. }
    \label{fig:feyn_SShh}
\end{figure}

The full set of `t-Hooft-Feynman gauge tree-level diagrams for $S^{(n_1)}S^{(n_2)} \rightarrow \varphi^{(n_3)}\varphi^{(n_4)}$ is depicted in Fig.~\ref{fig:feyn_SShh}, where we use the double line to indicate all three possible gravity intermediate states $h^{(i)}$, $A^{(i)}$, and $\varphi^{(i)}$. However, not all intermediate states contribute at $\mathcal{O}(s)$ in the high energy limit. To calculate the scattering amplitudes of scalar Goldstone bosons in the high energy limit, we only need to expand the Feynman rules to the leading order in momenta. Since each interaction term in the Lagrangian can contain at most two 4-derivatives $\partial_\mu$, the relevant non-vanishing Feynman rules at order $\mathcal{O}(E^2)$ are given in Appendix~\ref{sec:app_2}, where the vertices and terms below $\mathcal{O}(E^2)$ have been neglected: note that the contribution of the vector states $A^{(i)}$ is not relevant to this process at leading order.

Using the Feynman rules, we find the scattering amplitude at the leading order $\mathcal{O}(s)$ to be
\begin{equation}
\aligned
    \mathcal{M}\left[S^{(n_1)}S^{(n_2)} \rightarrow \varphi^{(n_3)}\varphi^{(n_4)}\right] =~& -\dfrac{\kappa^2s}{96}(3\cos2\theta+5)\sum_{i=0}^{\infty}\braket{k^{(n)}k^{(n)}f^{(i)}}\braket{f^{(i)}f_S^{(n)}f_S^{(n)}}\\
    &~+\dfrac{\kappa^2s}{12}\braket{k^{(n)}k^{(n)}f_S^{(n)}f_S^{(n)}} +\mathcal{O}(s^0)\\
    =~& \dfrac{\kappa^2s}{32}(1-\cos2\theta)\braket{k^{(n)}k^{(n)}f_S^{(n)}f_S^{(n)}}+\mathcal{O}(s^0),
\endaligned
\label{eq:ET_SShh}
\end{equation}
where we have abbreviated the overlap integrals  defining the mode couplings as \cite{Chivukula:2022kju}
\begin{equation}
    \braket{ f_1^{(n_1)}f_2^{(n_2)}\cdots} = \int_{z_1}^{z_2}dz~e^{3A(z)} f_1^{(n_1)}(z)f_2^{(n_2)}(z)\cdots.
    \label{eq:overlap-definition}
\end{equation}

This result agrees with the unitary gauge calculation given in Ref.~\cite{Chivukula:2023sua}, consistent with the Goldstone boson Equivalence theorem of Eq.~(\ref{eq:ge_1}). Note that the final amplitude at order $\mathcal{O}(s)$ can be written as proportional to the overlap integral of the product of the external state wave-functions $\braket{k^{(n)}k^{(n)}f_S^{(n)}f_S^{(n)}}$. This is because, at the leading order $\mathcal{O}(s)$, each interaction vertex must contain two 4-derivatives $\partial_\mu$ and have no KK mass dependence. Since the masses in the propagators can also be neglected in the high energy limit, one can always use the completeness relation to combine the two three-point overlap integrals into a four-point overlap integral. Note that the amplitude does not have any apparent vDVZ discontinuity, and therefore, one can safely take the $m_n\rightarrow0$ limit, which corresponds to the decoupling of the longitudinal mode.

\subsection{Scattering of two helicity-0 KK gravitons into two helicity-0 KK gravitons}

For the next example, we consider the scattering of two longitudinal KK gravitons,
\begin{equation}
    h^{(n_1)}_Lh^{(n_2)}_L \rightarrow h^{(n_3)}_{L}h^{(n_4)}_{L}.
\end{equation}
The Goldstone boson Equivalence theorem gives
\begin{equation}
    \mathcal{M}\left[h^{(n_1)}_Lh^{(n_2)}_L \rightarrow h^{(n_3)}_{L}h^{(n_4)}_{L}\right] = \mathcal{M}\left[\varphi^{(n_1)}\varphi^{(n_2)} \rightarrow \varphi^{(n_3)}\varphi^{(n_4)}\right] + \mathcal{O}(s^0).
\end{equation}
\begin{figure}
    \centering
    \includegraphics[width = 0.24\textwidth]{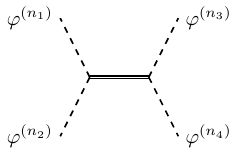}
    \includegraphics[width = 0.24\textwidth]{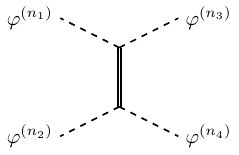}
    \includegraphics[width = 0.24\textwidth]{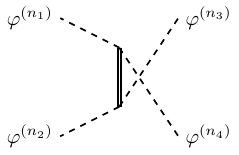}
    \includegraphics[width = 0.24\textwidth]{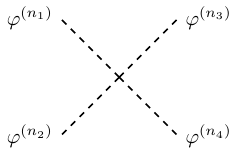}
    \caption{The Feynman diagrams for $\varphi^{(n_1)}\varphi^{(n_2)} \rightarrow \varphi^{(n_3)}\varphi^{(n_4)}$, where we use the double line to indicate all three possible intermediate states $h^{(i)}$, $A^{(i)}$, and $\varphi^{(i)}$. }
    \label{fig:feyn_hhhh}
\end{figure}
The Feynman diagrams for $\varphi^{(n_1)}\varphi^{(n_2)} \rightarrow \varphi^{(n_3)}\varphi^{(n_4)}$ are shown in Fig.~\ref{fig:feyn_hhhh}, where we use the double line to indicate all three possible intermediate states $h^{(i)}$, $A^{(i)}$, and $\varphi^{(i)}$. Based on the Feynman rules given in appendix~\ref{sec:app_2}, we find that the only diagrams that contribute at order $\mathcal{O}(s)$ are the $s$-, $t$- and $u$-channel diagrams with intermediate KK gravitons. And the scattering amplitude is then found to be
\begin{equation}
    \aligned
        \mathcal{M}\left[\varphi^{(n_1)}\varphi^{(n_2)} \rightarrow \varphi^{(n_3)}\varphi^{(n_4)}\right] =~& \dfrac{\kappa^2s}{16}(\sin^2\theta + 4\cot^2\dfrac{\theta}{2}+ 4\tan^2\dfrac{\theta}{2})\\
        &~\times \sum_{i=0}^{\infty}\braket{k^{(n)}k^{(n)}f^{(i)}}\braket{f^{(i)}k^{(n)}k^{(n)}} +\mathcal{O}(s^0)\\
        =~& \dfrac{\kappa^2s}{64}\dfrac{(\cos2\theta+7)^2}{\sin^2\theta}\braket{k^{(n)}k^{(n)}k^{(n)}k^{(n)}}+\mathcal{O}(s^0)
    \endaligned
    \label{eq:ET_hhhh}
    \end{equation}
This result agrees with the unitary gauge calculation given in Ref.~\cite{Chivukula:2020hvi}, once one uses the SUSY relation~\cite{Chivukula:2022kju},
\begin{equation}
    k^{(n)}(z) = -f^{(n)}(z)-\dfrac{2A'(z)}{m_n}g^{(n)}(z),
\end{equation}
to convert the graviton wave-functions in Ref.~\cite{Chivukula:2020hvi} into those of the scalar Goldstone bosons. As we explained in the previous subsection, the final amplitude is proportional to the overlap integral of the product of the external state wave-functions $\braket{k^{(n)}k^{(n)}k^{(n)}k^{(n)}}$, and no vDVZ discontinuity is present.

\medskip

We conclude this section by commenting on the relationship of our results to the ``double-copy" construction noted in \cite{PhysRevD.105.084005,Hang:2022rjp}. Motivated by the ``color-kinematic" duality relating gauge-theory and gravitational amplitudes~\cite{Bern:2008qj,Bern:2010ue,Bern:2019prr}, Hang and He note that since the massive spin-2 helicity-0 amplitudes grow only like ${\mathcal O}(s)$ and since these leading order amplitudes are KK mass-independent,  color-kinematic duality (which is exact in the massless theory) should also apply to the leading order in a compactified theory. Specifically, they demonstrate that an appropriate color-kinematic duality can be used to relate the high-energy scattering amplitude of the longitudinal modes of spin-1 KK bosons in a toroidally compactified five-dimensional gauge-theory to the high-energy scattering amplitudes of the helicity-0 modes of the corresponding spin-2 gravitational KK modes. Our result in Eq.~(\ref{eq:ET_hhhh}) above shows how their analysis generalizes to warped models: the kinematic factors remain precisely the same, but the couplings must be rescaled to account for the overlap integrals which give the mode-couplings of the (gauge- and gravitational) Goldstone bosons in the warped space.

While the results in this section demonstrate the validity of the Goldstone boson Equivalence theorems in Eqs.~(\ref{eq:ET_SShh}) and (\ref{eq:ET_hhhh}), we can actually compute full amplitudes using Eqs.~(\ref{eq:ward_A}) and (\ref{eq:ward_B}). We describe how to do so in the next section.

\section{A robust method of computing spin-2 scattering amplitudes}
\label{sec:exact}

\subsection{Method}
Studying the phenomenology of spin-2 gravitons requires the ability to reliably compute their scattering amplitudes. This is a challenge when working with models in warped geometries, where evaluating the exact tree-level scattering amplitudes in unitary gauge would technically require summing over an infinite number of intermediate KK states.

In practice, numerical computations of the helicity-0 spin-2 scattering amplitudes in unitary gauge are inherently unstable. The overlap integrals of the wave-functions can only be evaluated with finite precision, and one can only sum over a finite number of the intermediate KK states. These limitations introduce numerical errors which are amplified at high energies: the sum rules enforcing the cancellations~\cite{SekharChivukula:2019yul,Bonifacio:2019ioc,Chivukula:2020hvi} are only precisely true if one evaluates the overlap integrals exactly and sums over all possible intermediate states. These limitations of the numerical calculations reintroduce errors which in unitary gauge grow like ${\mathcal O}(s^5)$. Therefore, to evaluate amplitudes with sufficient accuracy at high energies in unitary gauge, one must not only evaluate all the overlap integrals with great precision but also sum over a large number of intermediate KK modes to keep the numerical errors under control~\cite{Chivukula:2020hvi}.

However, the earlier results in this paper enable us to mitigate those issues and achieve robust computation of spin-2 scattering amplitudes. Instead of using unitary gauge, one computes the amplitudes using `t-Hooft-Feynman gauge and applies the Ward identities described above to rewrite any matrix elements involving problematic external states as an appropriate combination of Goldstone boson amplitudes. 

This approach addresses each of the sources of bad high energy behavior.  On the one hand, all the internal propagators in the 't Hooft-Feynman gauge behave like $1/p^2$, eliminating the problematic high-energy behavior coming from unitary gauge projection operators in the propagators. On the other hand, the Ward identities in Eqs. ~(\ref{eq:ward_A}) and (\ref{eq:ward_B}) show that we can replace the matrix elements involving external helicity-0 and helicity-1 massive spin-2 states -- the states whose polarization tensors have potentially large high-energy behavior --  by a combination of amplitudes involving the corresponding Goldstone bosons and a residual ``spin-2" polarization vector ($\tilde{\epsilon}^{\mu\nu}$ in those equations) whose behavior at high-energies is mild.
Therefore, by combining these techniques, one can avoid the spurious high-energy growth which occurs in unitary gauge; the scattering amplitudes will converge as fast as the overlap integrals which determine the coupling among the various KK levels.

We will now illustrate this robust approach by applying it to analyze the full behavior of the scattering amplitudes described using the Equivalence theorem in section \ref{sec:example}.  First, we consider the scattering of two KK scalar bosons into a pair of KK gravitons.  Applying Eqs.(\ref{eq:ward_A}) and (\ref{eq:ward_B}) we find the amplitude can be written as
\begin{equation}
    \mathcal{M}\left[S^{(n_1)}S^{(n_2)} \rightarrow h^{(n_3)}_{\lambda_3}h^{(n_4)}_{\lambda_4}\right] = \sum_{\mathcal{F}_i=\phi,A,h}\left(\prod_{i=3}^4\zeta^*_{\lambda_i}(\mathcal{F}_{i})\right)\mathcal{M}\left[S^{(n_1)}S^{(n_2)}\rightarrow \mathcal{F}^{(n_3)}_{3,\lambda_3}\mathcal{F}^{(n_4)}_{4,\lambda_4} \right]
     \label{eq:exact_S}
\end{equation}
where the relative phases are defined as
\begin{equation}
    \zeta_\lambda(\phi) = \begin{cases} 1&\lambda=0\\0&{\rm else} \end{cases}, \qquad \zeta_\lambda(A) = \begin{cases} -i\sqrt{3}&\lambda=0\\-i&\lambda=\pm1\\0&{\rm else}\end{cases}, \qquad \zeta_\lambda(h) = 1.
\end{equation}
Each field $F_i$ on the right hand side of Eq.~(\ref{eq:exact_S}) represents $F_i=\phi,A,h$, where $\phi$ is the scalar Goldstone boson, $A^\mu_\lambda$ is the vector Goldstone boson with (unphysical) polarization $\tilde{\epsilon}^\mu_\lambda$, and $h^{\mu\nu}_\lambda$ is the KK graviton with (unphysical) polarization $\tilde{\epsilon}^{\mu\nu}_\lambda$,
\begin{equation}
    \tilde{\epsilon}^\mu_\lambda = \begin{cases}
        \epsilon^\mu_\lambda&\lambda = \pm\\
        \tilde{\epsilon}^\mu_0&\lambda = 0
    \end{cases},\qquad
    \tilde{\epsilon}^{\mu\nu}_\lambda = \begin{cases}
        \epsilon^{\mu\nu}_\lambda&\lambda = \pm2\\
        \tilde{\epsilon}^{\mu\nu}_\lambda&{\rm else}
    \end{cases}.
\end{equation}
Similarly, the scattering amplitude of two KK gravitons into a pair of KK gravitons is given by
\begin{equation}
    \mathcal{M}\left[h^{(n_1)}_{\lambda_1}h^{(n_2)}_{\lambda_2} \rightarrow h^{(n_3)}_{\lambda_3}h^{(n_4)}_{\lambda_4}\right] = \sum_{\mathcal{F}_i=\phi,A,h}\left(\prod_{i=1}^2\zeta(\mathcal{F}_{i,\lambda_i})\right)\left(\prod_{i=3}^4\zeta^*_{\lambda_i}(\mathcal{F}_{i})\right)\mathcal{M}\left[\mathcal{F}^{(n_1)}_{1,\lambda_1}\mathcal{F}^{(n_2)}_{2,\lambda_2} \rightarrow \mathcal{F}^{(n_3)}_{3,\lambda_3}\mathcal{F}^{(n_4)}_{4,\lambda_4}\right].
    \label{eq:exact_h}
\end{equation}
Again, we emphasize that the analyses leading to Eqs.~(\ref{eq:exact_S}) and (\ref{eq:exact_h})  hold for any background geometry, as long as the 't Hooft-Feynman in Eq.~(\ref{eq:GF}) exists and Eqs.(\ref{eq:ward_A}) and (\ref{eq:ward_B}) are true. However the details of the model can effect the couplings of the Goldstone bosons and therefore the Goldstone boson matrix elements themselves.

Next, we illustrate the numerical efficacy of using `t-Hooft-Feynman gauge and the Ward identities for these two amplitudes in RS1 where, due to the absence of the discrete momentum conservation present in toroidal models, the exact tree-level scattering amplitudes require summing over an infinite number of intermediate KK states. We set the numerical accuracy of our computation to be high (50 significant figures), to isolate and expose the issues arising from truncation error. The RS1 geometry is specified by the AdS curvature $k$ and ``compactification radius" $r_c$, where $k\pi r_c = \log(z_2/z_1)$, and the mode functions and overlap integrals depend only on the combination $kr_c$.

Following ~\cite{Chivukula:2020hvi}, we define the error due to truncating the sum over intermediate states at level $N$ by
\begin{equation}
    \Delta_{\rm trunc}(N,kr_c,s) = \left|\dfrac{\overline{\mathcal{M}^{[N]}}(kr_c,s)}{\overline{\mathcal{M}}(kr_c,s)} - 1\right|,
\end{equation}
where $\mathcal{M}^{[N]}$ is the scattering amplitude that only includes up to $N$ modes for intermediate KK states, and $\mathcal{M}$ is the exact scattering amplitude which we approximate using $\mathcal{M}\simeq\mathcal{M}^{[100]}$ computed via Eq.~(\ref{eq:exact_S}) or Eq.~(\ref{eq:exact_h}) in this work. In general, the scattering amplitudes could have different angular dependence at different truncation $N$ and different energies. To be representative, we average the scattering amplitudes over different values of the polar angle $\theta$~\footnote{While such average is a reasonable approach to estimate the errors on the cross section, it does not guarantee the accuracy of the angular distribution.},
\begin{equation}
    \overline{\mathcal{M}} = \dfrac{1}{9}\sum_{j=1}^9 \mathcal{M}(\theta = j \pi/10).
\end{equation}
Note that we have excluded the forward and backward region, $\theta < \pi/10$ and $\theta > 9\pi/10$, to avoid potential infrared divergences in the presence of massless intermediate particles in the $t$ and $u$ channels.

\begin{figure}[t]
\centering
\includegraphics[width=0.45\textwidth]{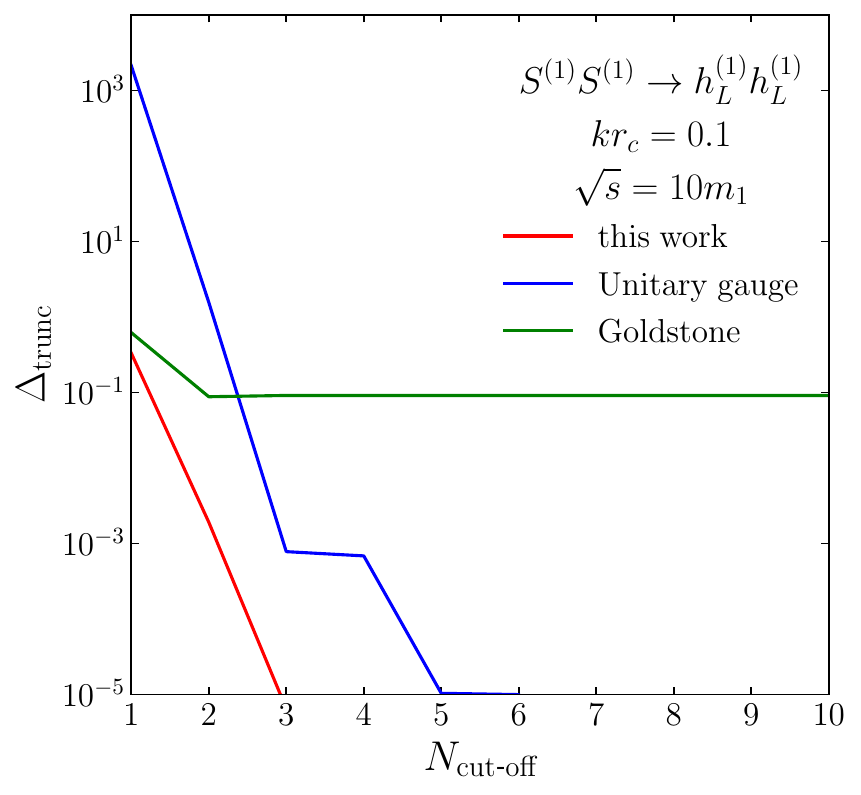}
\includegraphics[width=0.45\textwidth]{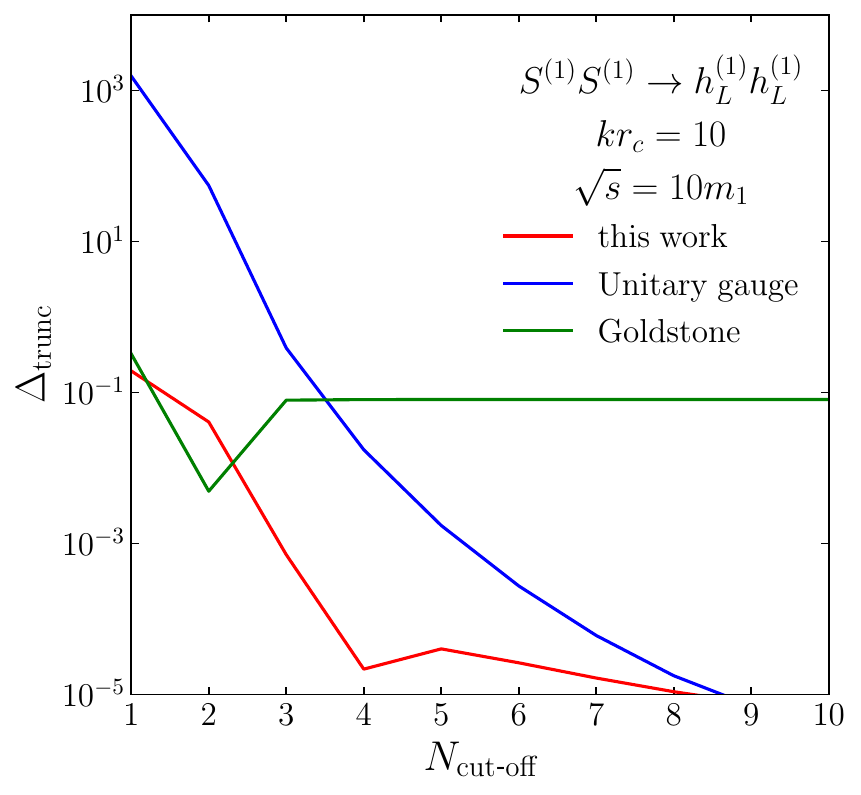}\\
\includegraphics[width=0.45\textwidth]{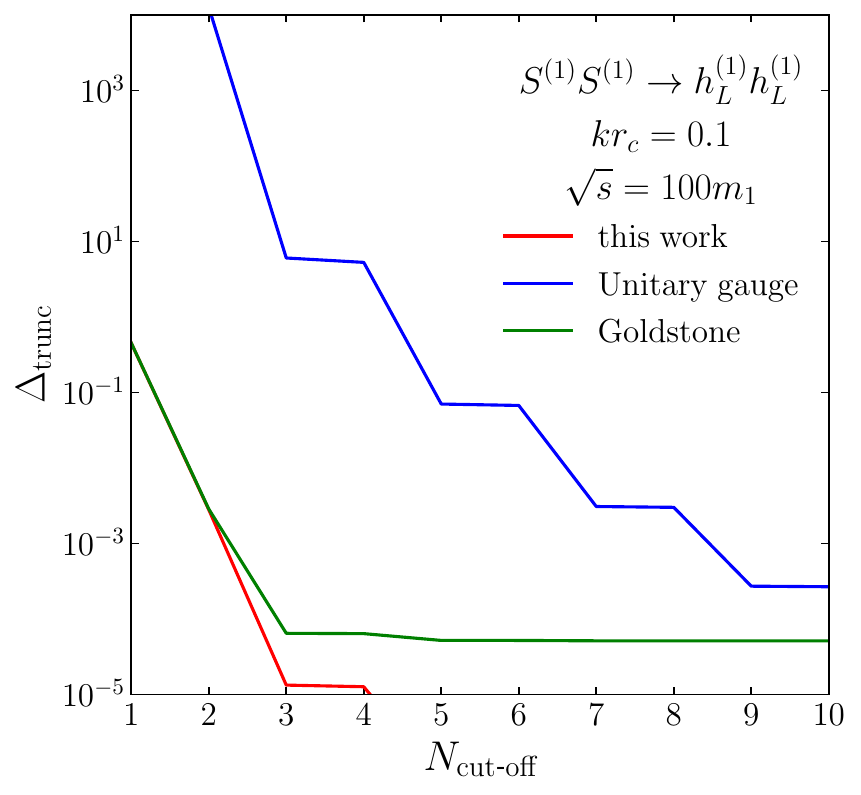}
\includegraphics[width=0.45\textwidth]{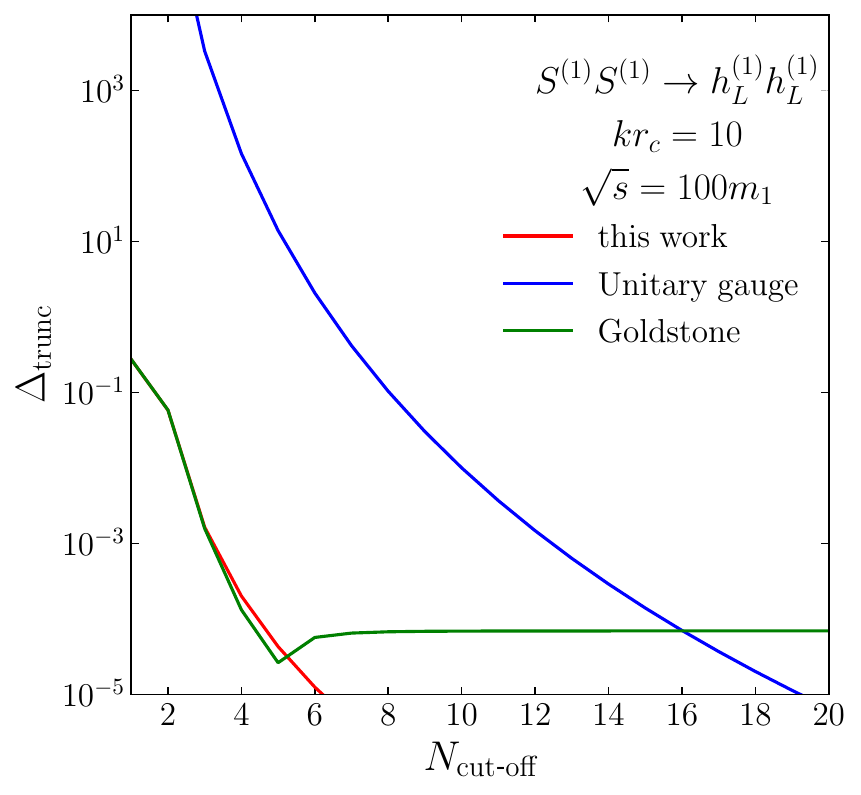}
\caption{The truncation error as a function of the number of included intermediate-state KK modes $N_{\rm cut\text{-}off}$ for the scattering of a pair of level-1 KK scalars in to a pair of longitudinal level-1 KK graviton $S^{(1)}S^{(1)}\rightarrow h^{(1)}_Lh^{(1)}_L$.  Outcomes are shown for benchmark models with different warping, $kr_c = 0.1$ (left column) and $kr_c = 10$ (right column), analyzed at two different scattering energies, $\sqrt{s} = 10m_1$ (upper panels) and $\sqrt{s} = 100m_1$ (lower panels). The results are computed via three methods: using our robust method involving `t-Hooft-Feynman gauge and the Ward identities as in  Eq.~(\ref{eq:exact_S}) (red), doing the calculation in unitary gauge (blue), and using the Goldstone boson Equivalence theorem to ${\mathcal O}(s)$ as in Eq.~(\ref{eq:ET_SShh})  (green).  Smaller truncation error implies a more reliable result for the scattering amplitude; lower $N_{\rm cut-off}$ implies a less resource-intensive computation. See text for a detailed discussion.}
\label{fig:DvN_SShh}
\end{figure}

\begin{figure}[t]
\centering
\includegraphics[width=0.45\textwidth]{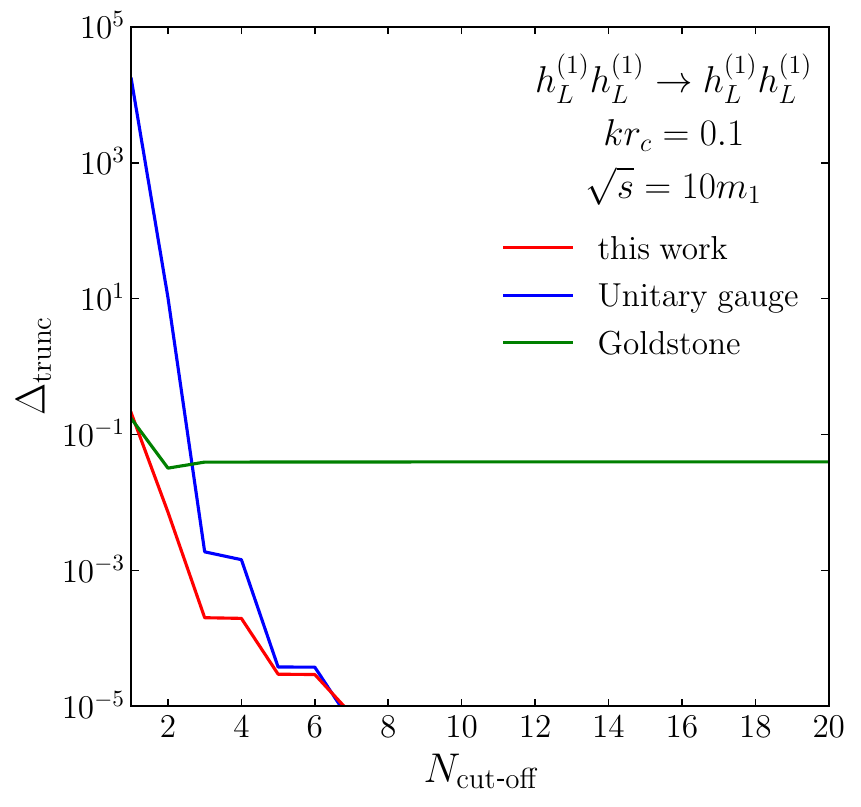}
\includegraphics[width=0.45\textwidth]{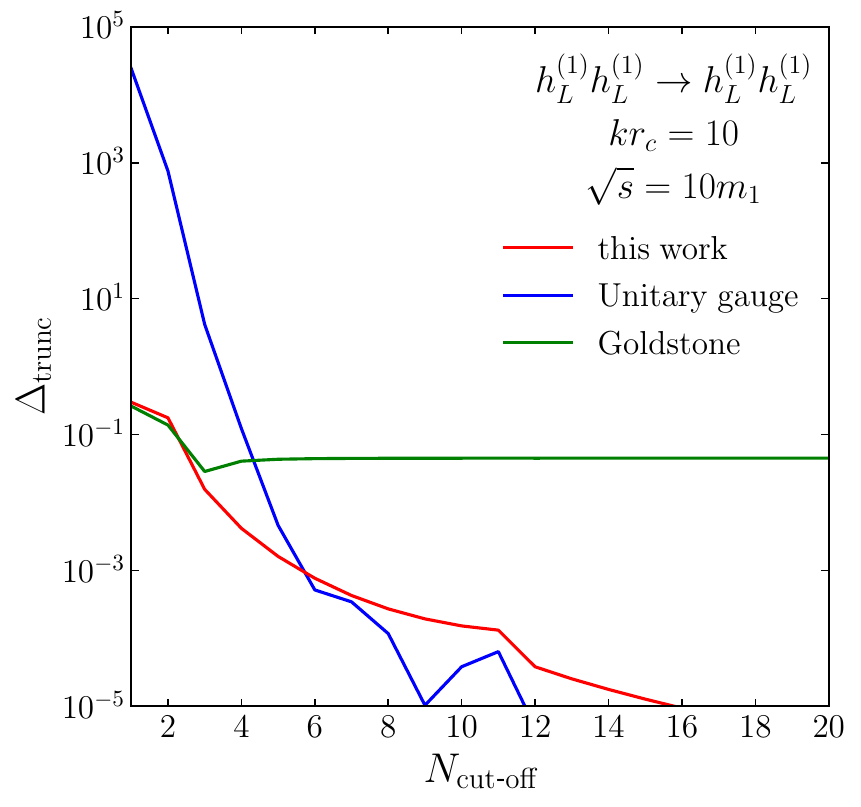}\\
\includegraphics[width=0.45\textwidth]{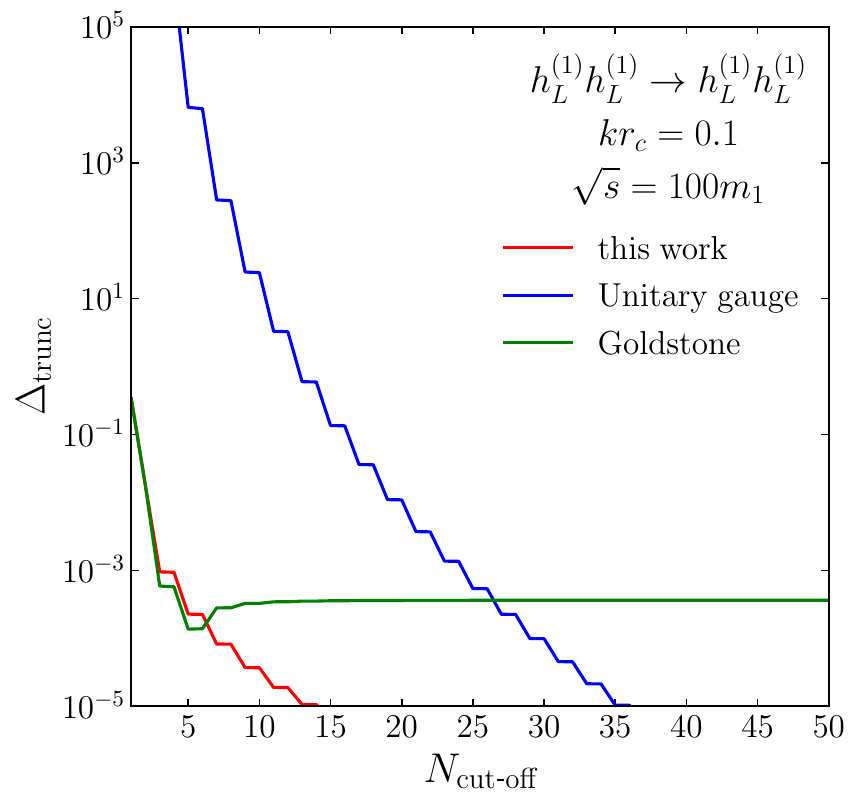}
\includegraphics[width=0.45\textwidth]{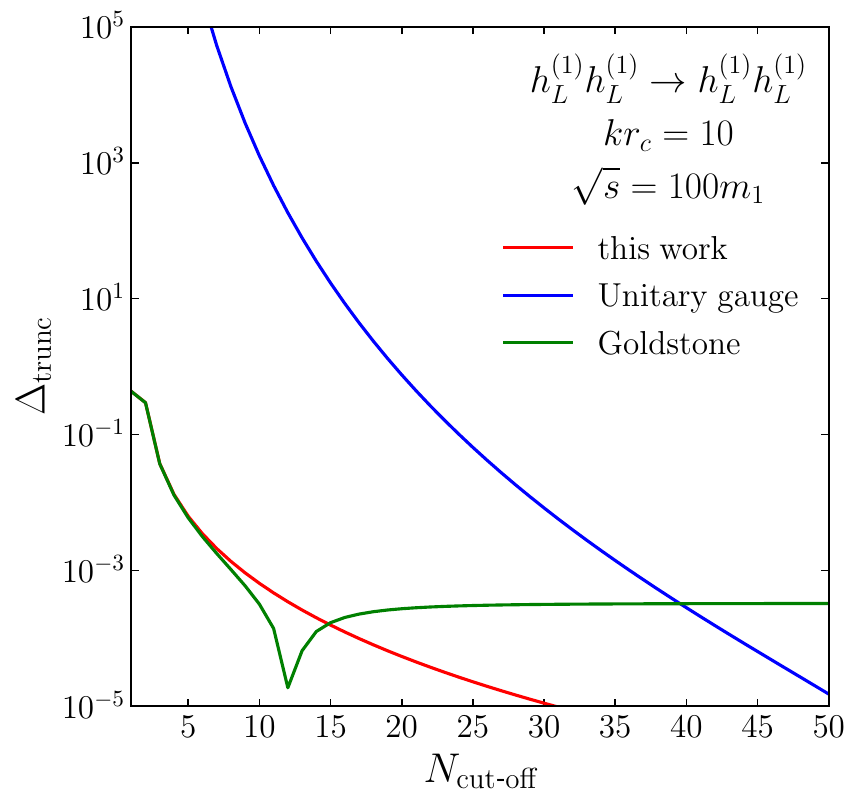}
\caption{The truncation error as a function of the number of included intermediate-state KK modes $N_{\rm cut\text{-}off}$ for the elastic scattering of a pair of longitudinal level-1 KK gravitons $h^{(1)}_Lh^{(1)}_L\rightarrow h^{(1)}_Lh^{(1)}_L$. The panel layout and curve labeling scheme are the same as in Fig.~\ref{fig:DvN_SShh}. See text for a detailed discussion.}
\label{fig:DvN_hhhh}
\end{figure}

\begin{figure}[t]
\centering
\includegraphics[width=0.45\textwidth]{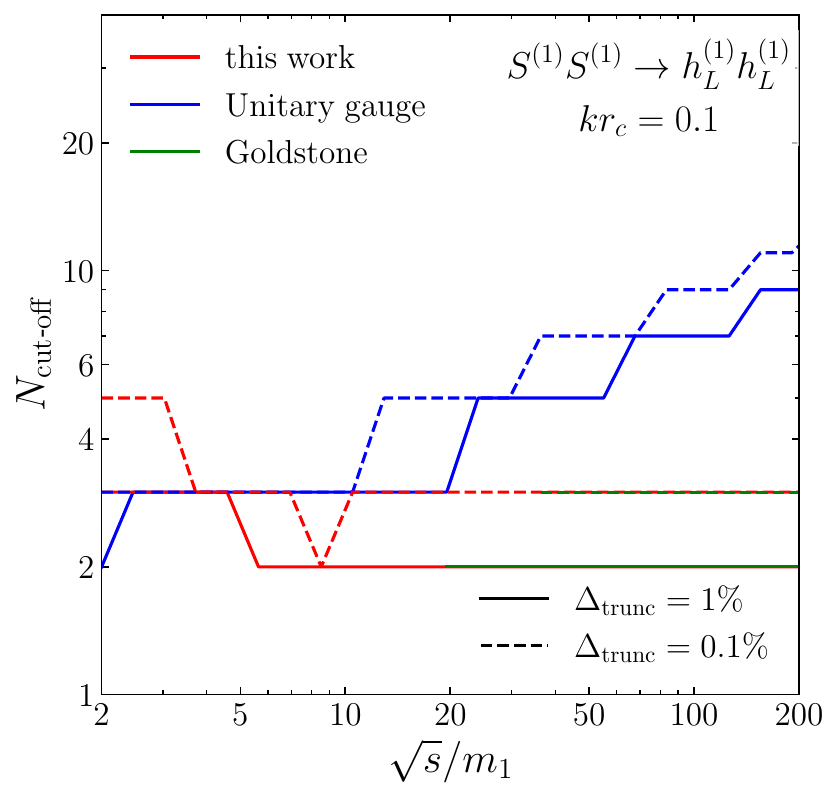}
\includegraphics[width=0.45\textwidth]{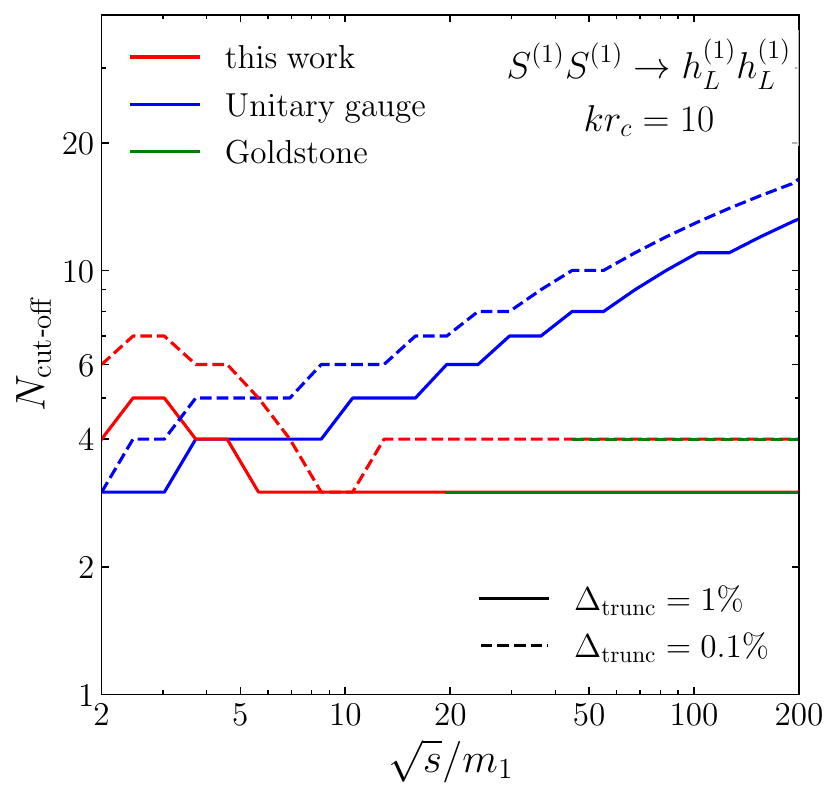}\\
\includegraphics[width=0.45\textwidth]{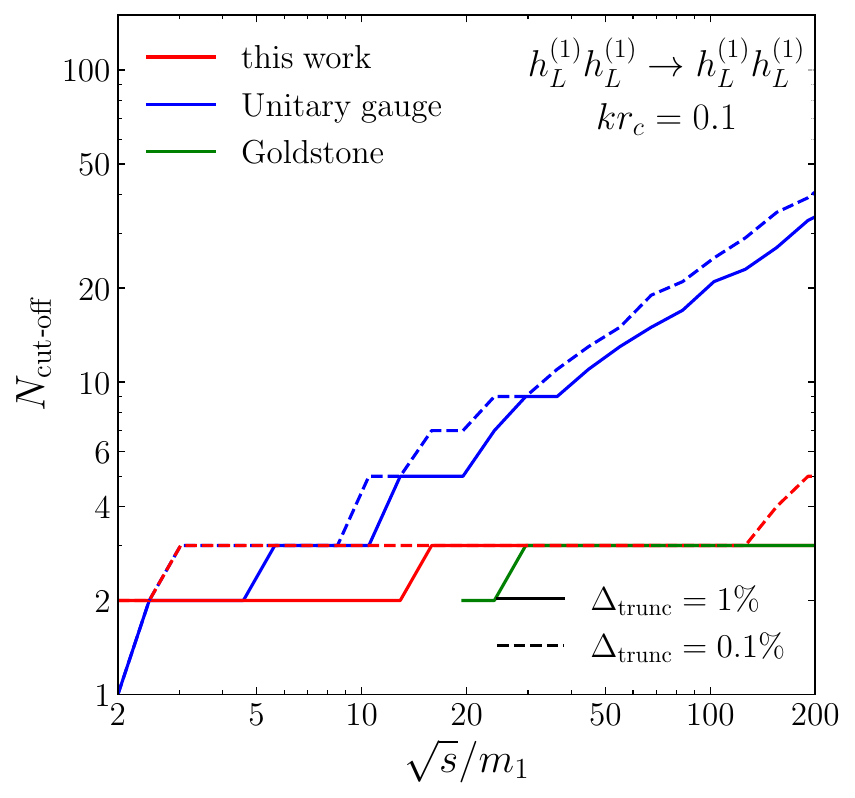}
\includegraphics[width=0.45\textwidth]{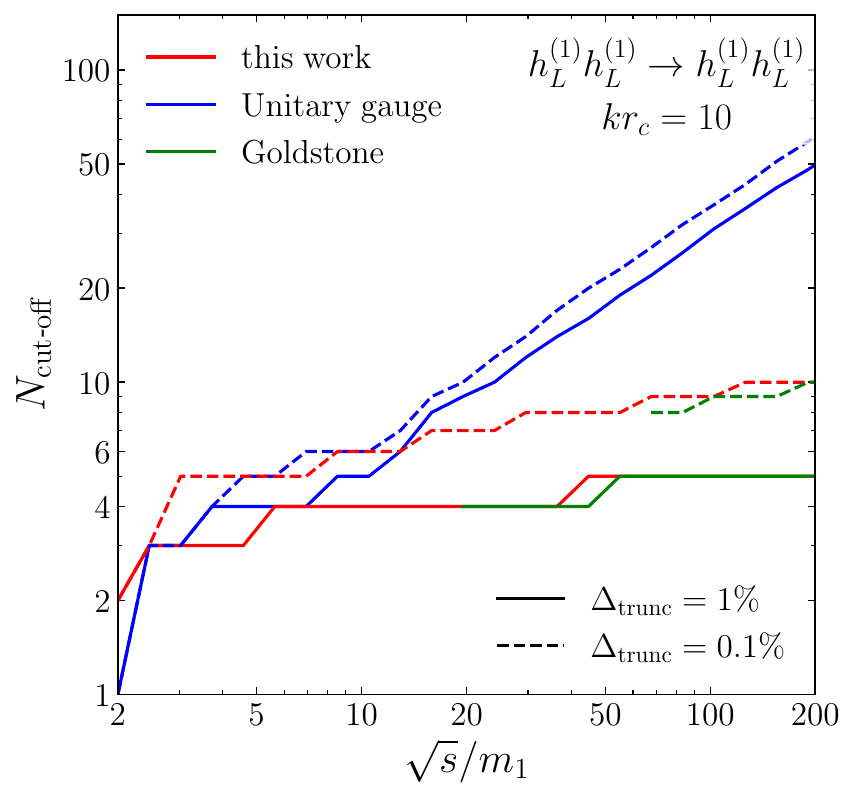}
\caption{The minimum number of intermediate KK modes one must include to achieve an accuracy of $\Delta_{\rm trunc} = 1\%$ (solid lines) or $\Delta_{\rm trunc} = 0.1\%$ (dashed lines), as a function of the ratio of the scattering energy over the mass. Upper panels are for $S^{(1)}S^{(1)}\rightarrow h^{(1)}_Lh^{(1)}_L$; lower panels are for $h^{(1)}_Lh^{(1)}_L\rightarrow h^{(1)}_Lh^{(1)}_L$. Outcomes are shown for benchmark models with different warping, $kr_c = 0.1$ (left column) and $kr_c = 10$ (right column) The curve labeling color scheme is the same as in Fig.~\ref{fig:DvN_SShh}.  See text for a detailed discussion.}
\label{fig:NvE}
\end{figure}

\subsection{Results}
We are now ready to discuss the results of our computations. Figures~\ref{fig:DvN_SShh} and \ref{fig:DvN_hhhh} display the relationship between truncation error and the number of included intermediate KK modes, while Figure~\ref{fig:NvE} illustrates the interplay between the number of intermediate states and the scattering energy in achieving a given level of computational precision.

Figure~\ref{fig:DvN_SShh} shows the truncation error as a function of the number of included KK modes $N_{\rm cut\text{-}off}$ for the scattering of a pair of level-1 KK scalars into a pair of longitudinal level-1 KK gravitons $S^{(1)}S^{(1)}\rightarrow h^{(1)}_Lh^{(1)}_L$. We have chosen two different benchmark models with different warping, $kr_c = 0.1$ (left column) and $kr_c = 10$ (right column), analyzed at two different scattering energies, $\sqrt{s} = 10m_1$ (upper panels) and $\sqrt{s} = 100m_1$ (lower panels). For simplicity, we have set the Lagrangian mass of the bulk scalar to be $M_S = 0$, such that the KK scalars and the gravitons have the same masses, $m_{S,n} = m_n$. The red lines are computed using the robust method from this paper involving `t-Hooft-Feynman gauge and the Ward identities, Eq.~(\ref{eq:exact_S}). This method clearly converges much faster as $N_{\rm cut\text{-}off}$ increases than computation in unitary gauge (blue lines), and the benefit is more pronounced for higher scattering energies (lower panels). For purposes of comparison, we also show (green lines) the results obtained when using the Goldstone boson Equivalence theorem in Eq.~(\ref{eq:ET_SShh}); this involves keeping only the ${\mathcal O}(s)$ contribution, so the resulting accuracy is no better than $m_1^2/s$ regardless of how many KK modes included. 

Fig.~\ref{fig:DvN_hhhh} displays the truncation errors for the elastic scattering of a pair of level-1 KK gravitons $h^{(1)}_Lh^{(1)}_L\rightarrow h^{(1)}_Lh^{(1)}_L$, which exhibit similar behaviors.

We note that, while the truncation error does converge as fast as the overlap integrals when one uses 't-Hooft-Feynman gauge and the Ward identities, different overlap integrals may converge at different rates. Such variance in convergence can lead to cases at intermediate scattering energies (upper panels) and small truncation error values (i.e., large enough $N_{\rm cut-off}$) where the unitary gauge computation yields a truncation error equal to or smaller than the result of our robust method. For example, as shown in upper right panel in Fig.~\ref{fig:DvN_hhhh}, the unitary gauge result (blue) and the 't-Hooft-Feynman gauge result (red) intersect at $N_{\text{cut-off}}\sim 6$, and the unitary gauge computation errors drop faster after that point when including more intermediate states. One should notice, however, that the errors in both cases are already less than 0.1\%. In this case, the truncation errors are expected to have the scaling behaviors of
\begin{equation}
    \Delta_{\rm trunc}\propto \begin{cases}
    \braket{f^{(1)}f^{(1)}f^{(N_{\text{cut-off}})}}^2\times \left(\dfrac{s}{m_1^2}\right)^4, \qquad \text{unitary gauge}, \\
    \max\limits_{f_i \in \{f, g, k\}}\braket{f_1^{(1)}f_2^{(1)}f_3^{(N_{\text{cut-off}})}}^2, \qquad \text{'t-Hooft-Feynman gauge}.
    \end{cases}
\end{equation}
We find that numerically some of the overlap integrals in the 't-Hooft-Feynman gauge computation, for example $\braket{k^{(1)}k^{(1)}f^{(N_{\text{cut-off}})}}$, converge more slowly than $\braket{f^{(1)}f^{(1)}f^{(N_{\text{cut-off}})}}$ (the only overlap needed in unitary gauge) as $N_{\text{cut-off}}$ increases. Therefore, for small enough truncation errors, such difference in convergence could overcome the amplification factor $s^4/m_1^8$ in the unitary gauge computation, which causes the intersection of the blue and red lines in Fig.~\ref{fig:DvN_hhhh} at $N_{\text{cut-off}}\sim6$ at intermediate energies. 

As the scattering energy increases, however, the unitary gauge calculation always requires more intermediate states to reach an accurate result. Figure~\ref{fig:NvE} shows the minimum number of intermediate KK modes one has to include to achieve the accuracy of $\Delta_{\rm trunc} = 1\%$ (solid lines) or $\Delta_{\rm trunc} = 0.1\%$ (dashed lines), as a function of the ratio of the scattering energy over the mass. The upper panels are plotted for the scattering of $S^{(1)}S^{(1)}\rightarrow h^{(1)}_Lh^{(1)}_L$, and the lower panels are plotted for the scattering of $h^{(1)}_Lh^{(1)}_L\rightarrow h^{(1)}_Lh^{(1)}_L$, with $kr_c = 0.1$ (left column) and $kr_c = 10$ (right panels). As shown by the red lines computed using our robust method (Eq.~(\ref{eq:exact_S}) and (\ref{eq:exact_h})), the number of included intermediate KK modes required to achieve a given accuracy stays constant at higher energies. On the contrary, if one employs unitary gauge (blue lines), the number of intermediate KK modes that must be included to achieve a given accuracy grows dramatically with a power law behavior as the energy increases, due to the bad high-energy behavior of the truncated unitary-gauge amplitudes. The red and blue lines are comparable only at lower energies where the residual contribution of the $\mathcal{O}(s^\sigma)$, for $\sigma >1$, caused by the truncation, are less important. 
For comparison, we also show the Goldstone boson Equivalence theorem results in green; note that this method can reach 1\% accuracy only at very high energies $\sqrt{s}\gtrsim 20 m_1$, so the solid green lines do not extend to lower energies in these plots and a dashed green line appears only at high energy for large $kr_c$.

\section{Conclusions}
\label{sec:conclusion}

In this work, extending results by Hang and He, we have showed how the residual five-dimensional diffeomorphism symmetries of compactified gravitational theories with a warped extra dimension imply Equivalence theorems which ensure that the scattering amplitudes of helicity-0 and helicity-1 spin-2 Kaluza-Klein states equal (to leading order in scattering energy) those of the corresponding Goldstone bosons present in the `t-Hooft-Feynman gauge. We explicitly calculated these amplitudes in terms of the Goldstone bosons in the Randall-Sundrum model and checked the correspondence to previous unitary-gauge computations. We also introduced a novel and robust method for accurately computing amplitudes for scattering of the spin-2 states both among themselves and with matter -- and demonstrated that this method outperforms unitary gauge calculations especially at higher scattering energies.

It is interesting to consider how these results generalize to 
other background geometries. As we mention in Sec.~\ref{sec:longitudinal}, since such gauge fixing conditions have the exactly same form in toroidal compactification \cite{PhysRevD.105.084005,Hang:2022rjp}, RS1 \cite{Lim:2007fy,Lim:2008hi}, or GW models \cite{Chivukula:2022kju}, the Ward identities themselves will be of the same {\it form} in all cases regardless of the background geometry. The fact that the scattering amplitudes of the helicity-0 and helicity-1 states can be written in the form of 
Eqs.~(\ref{eq:ward_A}) and (\ref{eq:ward_B}), and that the 
scattering amplitudes of the helicity-0 and helicity-1 states of the KK-gravitons equal those of the corresponding Goldstone bosons in `t-Hooft-Feynman gauge is also true in all of these cases. Therefore the power-counting arguments given, showing that two-to-two helicity-0 spin-2 scattering processes cannot grow any faster than ${\mathcal O}(s)$, are also generally true.

However, to use the Equivalence theorem to compute scattering amplitudes one needs to be able to explicitly compute the corresponding `t-Hooft-Feynman gauge Goldstone boson couplings.
In general, this depends on the details of the model in at least two ways. First, the mode-functions of the KK states (the analogs of $f^{(n)}(z)$, $g^{(n)}(z)$, and $k^{(n)}(z)$ in RS1) will be different in different background geometries and must be computed.
Second, as in the case of the Goldberger-Wise model~\cite{Goldberger:1999uk,Goldberger:1999un,Chivukula:2022kju}, the scalar Goldstone bosons may be linear combinations of gravitational and bulk scalar fields, and their interactions will therefore also be dependent on the bulk scalar potential.\footnote{This is analogous to using the Equivalence theorem for longitudinal $W$-boson scattering in the Higgs-doublet standard model: the Goldstone boson scattering amplitudes include Higgs-exchange contributions, and are not determined by the gauge theory alone.} \looseness=-1

Finally, we note that since the analyses presented here rely on the residual background 5D diffoemorphism symmetries of the theory, it should be possible to extend these results to consider processes beyond the tree-level analyses given here.\looseness=-1

\noindent {\bf Acknowledgements:} The authors thank Dennis Foren for collaboration during the initial stages of this work. The work of RSC, EHS, and XW was supported in part by the National Science Foundation under Grant No.~PHY-2210177. JAG acknowledges the support he has
received for his research through the provision of an Australian Government Research Training Program Scholarship.
Support for this work was provided by the University of Adelaide and the Australian Research Council through the Centre of Excellence for Dark Matter Particle Physics (CE200100008). The work of KM was supported in part by the National Science Foundation under Grant No.~PHY-2310497. JAG and DS thank Anthony G. Williams for fruitful discussions. \looseness=-1

\pagebreak

\appendix
\section{Notation}
\label{sec:app_1}

The 5D Lagrangian of the RS1 model is given by 
\begin{equation}
    \mathcal{L} =\mathcal{L}_{EH} + \mathcal{L}_{\rm CC} + \Delta\mathcal{L} + \mathcal{L}_{\rm GF} + \mathcal{L}_{\rm m}
\end{equation}
where $\mathcal{L}_{EH}$ and $\mathcal{L}_{\rm CC}$ are the usual Einstein-Hilbert and cosmological constant terms respectively. The $\Delta\mathcal{L}$ term is a total derivative term required for a well defined variational principle for the action. $\mathcal{L}_{\rm GF}$ is gauge fixing term and $\mathcal{L}_{\rm m}$ is the Lagrangian of the matter fields\footnote{The mathematical expressions for each of these terms are provided in \cite{Chivukula:2020hvi} and in conformal co-ordinates in \cite{Chivukula:2022kju} }.
The RS1 line element in conformal coordinates  $(x_{\mu},z)$ is written as, 
\begin{equation}
    ds^2 = e^{2A(z)}(\eta_{\mu\nu}dx^\mu dx^\nu - dz^2),
\end{equation}
where the background 4D Minkowski metric $\eta_{\mu\nu}\equiv{\rm diag}(+1,-1,-1,-1)$ is used to raise and lower indices, and $z$ lies in the interval $z_1=1/k \le z \le z_2=e^{k\pi r_c}/k$, where $k$ is the AdS curvature and $r_c$ is the RS1 ``compactification radius," a measure of the size of the internal dimension.\footnote{The dimensionless ratio $kr_c$ is a convenient measure of how ``warped" the internal space is and determines the mode functions and overlap integrals needed to compute scattering amplitudes.}
The warp factor $A(z)$ is given by, 
\begin{equation}
   A(z) = -\ln(kz)~,
\end{equation}

The metric, including fluctuations, can then be written as, 
\begin{equation}
    G_{MN} = e^{2A(z)}\begin{pmatrix}
        e^{-\kappa\varphi/\sqrt{6}}(\eta_{\mu\nu}+\kappa h_{\mu\nu}(x,z)) & \frac{\kappa}{\sqrt{2}}A_\mu(x,z) \\
        \frac{\kappa}{\sqrt{2}}A_\mu(x,z) & -\left(1+\frac{\kappa}{\sqrt{6}}\varphi(x,z)\right)^2
    \end{pmatrix}~,
    \label{eq:background-metric}
\end{equation}
The metric fluctuations $h_{\mu\nu}(x,z)$ define the spin-2 fluctuations in 4D, while  ${A}_\mu(x,z)$
and ${\varphi}(x,z)$ are the spin-1 and spin-0 fluctuations respectively.
Here we have followed the notation in Ref.~\cite{Chivukula:2022kju,Lim:2007fy,Lim:2008hi} and we refer the reader there for details.

\section{Feynman Rules}
\label{sec:app_2}
The relevant Feynman rules that are used in Sec.~\ref{sec:example} are given below, 
\begin{flalign}
        \begin{gathered}\includegraphics[width = 0.2\textwidth]{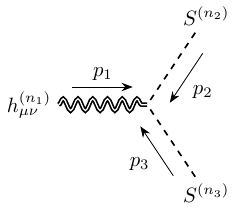}\end{gathered} & \begin{gathered}=~ \dfrac{i\kappa}{2}\braket{f^{(n_1)}f_S^{(n_2)}f_S^{(n_3)}}\left(p_2^\mu p_3^\nu + p_2^\nu p_3^\mu-\eta^{\mu\nu}p_2\cdot p_3\right) + \mathcal{O}\left((p_i)^0\right)\end{gathered}, \\
        \begin{gathered}\includegraphics[width = 0.2\textwidth]{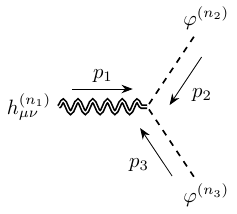}\end{gathered} & \begin{gathered}=~\dfrac{i\kappa}{24}\braket{f^{(n_1)}k^{(n_2)}k^{(n_3)}}\left[4p_1^\mu p_1^\nu - 12p_2^\mu p_2^\nu - 12p_3^\mu p_3^\nu\right.\\
        \left.-\eta^{\mu\nu}\left(2p_1^2+6p_2^2+6p_3^2\right)\right] + \mathcal{O}\left((p_i)^0\right)\end{gathered},\\
        \begin{gathered}\includegraphics[width = 0.2\textwidth]{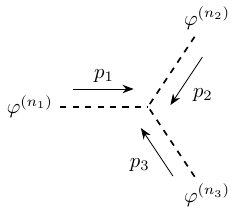}\end{gathered} & \begin{gathered}=~-\dfrac{i\kappa}{\sqrt{6}}\braket{k^{(n_1)}k^{(n_2)}k^{(n_3)}}\left(p_1^2+p_2^2+p_3^2\right) + \mathcal{O}\left((p_i)^0\right)\end{gathered},\\
        \begin{gathered}\includegraphics[width = 0.2\textwidth]{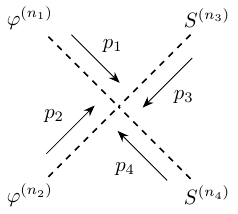}\end{gathered} & \begin{gathered}=~\dfrac{i\kappa^2}{6}\braket{k^{(n_1)}k^{(n_2)}f_S^{(n_3)}f_S^{(n_4)}}\left(p_3\cdot p_4\right) + \mathcal{O}\left((p_i)^0\right)\end{gathered},\\
        \begin{gathered}\includegraphics[width = 0.2\textwidth]{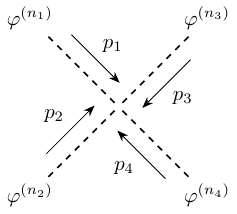}\end{gathered} & \begin{gathered}=~\dfrac{i\kappa^2}{4}\braket{k^{(n_1)}k^{(n_2)}k^{(n_3)}k^{(n_4)}}\left(\sum_{i=1}^4 p_i^2\right) + \mathcal{O}\left((p_i)^0\right).\end{gathered}
\end{flalign}

\section{Symmetry algebra of the residual RS1 5D diffeomorphisms}

\label{sec:app_3}

Following the exposition of Duff and Dolan in Ref.~\cite{Dolan:1983aa}, we can explicitly identify the residual discrete 5D diffeomorphism symmetries  that are preserved in our compactified warped model.
The infinitesimal coordinate transformation parameter $\xi^M$ ($\delta x^\mu=\xi^\mu$ and $\delta z= \xi^5$) can be expanded,
\begin{eqnarray}
    \xi^\mu(x,z) &=& \sum_nf^{(n)}(z)\ \xi^\mu_n(x),\label{eq:coordinate1}\\
    \xi^5(x,z) &=& \sum_ng^{(n)}(z)\ \xi^5_n(x), \label{eq:coordinate2}
\end{eqnarray}
where the functions $f^{(n)}(z)$ and $g^{(n)}(z)$ are precisely the eigenfunctions that appear in the expansions of the KK tensor and vector modes in Eqs.~(\ref{eq:KK_1u}) and (\ref{eq:KK_2u}). Note that the functions $g^{(n)}(z)$ start with $n=1$ and are chosen to vanish at the boundaries ($\xi^M(z_{1,2})=0$) of RS1 so that the coordinate transformations do not change the location of the branes.

As emphasized in \cite{Dolan:1983aa}, ordinary 5D general coordinate transformations in flat space can be regarded as the local gauge transformations that correspond to the corresponding global Poincare algebra, 
\begin{eqnarray}
    \xi^\mu_n &=& a_n{}^\mu + \omega_n{}^\mu{}_\nu x^\nu,\\
    \xi^5_n &=& c_n.
\end{eqnarray}
Therefore, the generators corresponding to the residual transformations in Eqs.~(\ref{eq:coordinate1}) and (\ref{eq:coordinate2}) are
\begin{eqnarray}
    P^\mu_n &=& f^{(n)}\partial^\mu,\\
    M^{\mu\nu}_n &=& f^{(n)}(x^\nu\partial^\mu-x^\mu\partial^\nu),\\
    Q_n &=& g^{(n)}\partial_z.
\end{eqnarray}
These generators define the following infinite parameter Lie algebra
\begin{eqnarray}
    \left[P^\mu_n,P^\nu_m\right] &=& 0,\\
    \left[M^{\mu\nu}_n,P^\sigma_m\right] &=& \sum_l\braket{f^{(n)}f^{(m)}f^{(l)}}\left(\eta^{\mu\sigma}P^\nu_l-\eta^{\nu\sigma}P^\mu_l\right),\\
    \left[M^{\mu\nu}_n,M^{\rho\sigma}_m\right] &=& \sum_l\braket{f^{(n)}f^{(m)}f^{(l)}}\left(\eta^{\mu\rho}M^{\nu\sigma}_l+\eta^{\nu\sigma}M^{\mu\rho}_l-\eta^{\mu\sigma}M^{\nu\rho}_l-\eta^{\nu\rho}M^{\mu\sigma}_l\right),\\
    \left[Q_n,Q_m\right] &=& \sum_l\left[\dfrac{m_m-m_n}{2}\left(\braket{f^{(m)}g^{(n)}g^{(l)}}+\braket{g^{(m)}f^{(n)}g^{(l)}}\right) \right.\\
    && \left.+ \dfrac{m_m+m_n}{2}\left(\braket{f^{(m)}g^{(n)}g^{(l)}}-\braket{g^{(m)}f^{(n)}g^{(l)}}\right)\right]Q_l,\\
    \left[Q_n,P^\mu_m\right] &=& \sum_lm_m\braket{g^{(m)}g^{(n)}f^{(l)}}P^\mu_l,\\
    \left[Q_n,M^{\mu\nu}_m\right] &=& \sum_lm_m\braket{g^{(m)}g^{(n)}f^{(l)}}M^{\mu\nu}_l.
\end{eqnarray}
In the last expressions, which require evaluating derivatives of the mode functions, we use the SUSY structure of the mode eigenequations \cite{Lim:2007fy,Lim:2008hi,Chivukula:2022kju}.
Note that the masses of the eigenmodes $m_n$ and the overlap integrals defined by Eq.~(\ref{eq:overlap-definition}) appear in the structure-constants of this algebra.
As in the case of toroidal compactification \cite{Dolan:1983aa}, the symmetries with $n\ge 1$ are spontaneously broken giving rise to the (space-time) Goldstone bosons $A^{(n)}_\mu$ and $\varphi^{(n)}$ which are ``eaten" by the corresponding spin-2 modes. $h^{(0)}_{\mu\nu}$ is the massless 4D graviton, and there is no broken symmetry corresponding to the radion, $\varphi^{(0)}$. The radion can be given a mass via the Goldberger-Wise mechanism \cite{Goldberger:1999uk,Goldberger:1999un} while still respecting a residual 5D diffeomorphism invariance \cite{Chivukula:2022kju}.

In the case of toroidal compactification, where the internal wavefunctions are simple trigonometric functions and which has a discrete momentum conservation corresponding to discrete global translations in the extra dimension, the above algebra reduces to the Kac-Moody algebra in Ref.~\cite{Dolan:1983aa}.

\newpage
\bibliographystyle{unsrt}   
\bibliographystyle{apsrev4-1.bst}
\bibliography{main}

\end{document}